\makeatletter

\font\tenmsx=msxm10
\font\sevenmsx=msxm7
\font\fivemsx=msxm5
\font\tenmsy=msym10
\font\sevenmsy=msym7
\font\fivemsy=msym5
\newfam\msxfam
\newfam\msyfam
\textfont\msxfam=\tenmsx  \scriptfont\msxfam=\sevenmsx
  \scriptscriptfont\msxfam=\fivemsx
\textfont\msyfam=\tenmsy  \scriptfont\msyfam=\sevenmsy
  \scriptscriptfont\msyfam=\fivemsy

\def\hexnumber@#1{\ifnum#1<10 \number#1\else
 \ifnum#1=10 A\else\ifnum#1=11 B\else\ifnum#1=12 C\else
 \ifnum#1=13 D\else\ifnum#1=14 E\else\ifnum#1=15 F\fi\fi\fi\fi\fi\fi\fi}

\def\msx@{\hexnumber@\msxfam}
\def\msy@{\hexnumber@\msyfam}
\mathchardef\boxdot="2\msx@00
\mathchardef\boxplus="2\msx@01
\mathchardef\boxtimes="2\msx@02
\mathchardef\square="0\msx@03
\mathchardef\blacksquare="0\msx@04
\mathchardef\centerdot="2\msx@05
\mathchardef\lozenge="0\msx@06
\mathchardef\blacklozenge="0\msx@07
\mathchardef\circlearrowright="3\msx@08
\mathchardef\circlearrowleft="3\msx@09
\mathchardef\rightleftharpoons="3\msx@0A
\mathchardef\leftrightharpoons="3\msx@0B
\mathchardef\boxminus="2\msx@0C
\mathchardef\Vdash="3\msx@0D
\mathchardef\Vvdash="3\msx@0E
\mathchardef\vDash="3\msx@0F
\mathchardef\twoheadrightarrow="3\msx@10
\mathchardef\twoheadleftarrow="3\msx@11
\mathchardef\leftleftarrows="3\msx@12
\mathchardef\rightrightarrows="3\msx@13
\mathchardef\upuparrows="3\msx@14
\mathchardef\downdownarrows="3\msx@15
\mathchardef\upharpoonright="3\msx@16

\mathchardef\downharpoonright="3\msx@17
\mathchardef\upharpoonleft="3\msx@18
\mathchardef\downharpoonleft="3\msx@19
\mathchardef\rightarrowtail="3\msx@1A
\mathchardef\leftarrowtail="3\msx@1B
\mathchardef\leftrightarrows="3\msx@1C
\mathchardef\rightleftarrows="3\msx@1D
\mathchardef\Lsh="3\msx@1E
\mathchardef\Rsh="3\msx@1F
\mathchardef\rightsquigarrow="3\msx@20
\mathchardef\leftrightsquigarrow="3\msx@21
\mathchardef\looparrowleft="3\msx@22
\mathchardef\looparrowright="3\msx@23
\mathchardef\circeq="3\msx@24
\mathchardef\succsim="3\msx@25
\mathchardef\gtrsim="3\msx@26
\mathchardef\gtrapprox="3\msx@27
\mathchardef\multimap="3\msx@28
\mathchardef\therefore="3\msx@29
\mathchardef\because="3\msx@2A
\mathchardef\doteqdot="3\msx@2B

\mathchardef\triangleq="3\msx@2C
\mathchardef\precsim="3\msx@2D
\mathchardef\lesssim="3\msx@2E
\mathchardef\lessapprox="3\msx@2F
\mathchardef\eqslantless="3\msx@30
\mathchardef\eqslantgtr="3\msx@31
\mathchardef\curlyeqprec="3\msx@32
\mathchardef\curlyeqsucc="3\msx@33
\mathchardef\preccurlyeq="3\msx@34
\mathchardef\leqq="3\msx@35
\mathchardef\leqslant="3\msx@36
\mathchardef\lessgtr="3\msx@37
\mathchardef\backprime="0\msx@38
\mathchardef\risingdotseq="3\msx@3A
\mathchardef\fallingdotseq="3\msx@3B
\mathchardef\succcurlyeq="3\msx@3C
\mathchardef\geqq="3\msx@3D
\mathchardef\geqslant="3\msx@3E
\mathchardef\gtrless="3\msx@3F
\mathchardef\sqsubset="3\msx@40
\mathchardef\sqsupset="3\msx@41
\mathchardef\trianglerighteq="3\msx@44
\mathchardef\trianglelefteq="3\msx@45
\mathchardef\bigstar="0\msx@46
\mathchardef\between="3\msx@47
\mathchardef\blacktriangledown="0\msx@48
\mathchardef\blacktriangleright="3\msx@49
\mathchardef\blacktriangleleft="3\msx@4A
\mathchardef\blacktriangle="0\msx@4E
\mathchardef\triangledown="0\msx@4F
\mathchardef\eqcirc="3\msx@50
\mathchardef\lesseqgtr="3\msx@51
\mathchardef\gtreqless="3\msx@52
\mathchardef\lesseqqgtr="3\msx@53
\mathchardef\gtreqqless="3\msx@54
\mathchardef\Rrightarrow="3\msx@56
\mathchardef\Lleftarrow="3\msx@57
\mathchardef\veebar="2\msx@59
\mathchardef\barwedge="2\msx@5A
\mathchardef\doublebarwedge="2\msx@5B
\mathchardef\angle="0\msx@5C
\mathchardef\measuredangle="0\msx@5D
\mathchardef\sphericalangle="0\msx@5E
\mathchardef\varpropto="3\msx@5F
\mathchardef\smallsmile="3\msx@60
\mathchardef\smallfrown="3\msx@61
\mathchardef\Subset="3\msx@62
\mathchardef\Supset="3\msx@63
\mathchardef\Cup="2\msx@64

\mathchardef\Cap="2\msx@65

\mathchardef\curlywedge="2\msx@66
\mathchardef\curlyvee="2\msx@67
\mathchardef\leftthreetimes="2\msx@68
\mathchardef\rightthreetimes="2\msx@69
\mathchardef\subseteqq="3\msx@6A
\mathchardef\supseteqq="3\msx@6B
\mathchardef\bumpeq="3\msx@6C
\mathchardef\Bumpeq="3\msx@6D
\mathchardef\lll="3\msx@6E

\mathchardef\ggg="3\msx@6F

\mathchardef\circledS="0\msx@73
\mathchardef\pitchfork="3\msx@74
\mathchardef\dotplus="2\msx@75
\mathchardef\backsim="3\msx@76
\mathchardef\backsimeq="3\msx@77
\mathchardef\complement="0\msx@7B
\mathchardef\intercal="2\msx@7C
\mathchardef\circledcirc="2\msx@7D
\mathchardef\circledast="2\msx@7E
\mathchardef\circleddash="2\msx@7F
\def\ulcorner{\delimiter"4\msx@70\msx@70 }
\def\urcorner{\delimiter"5\msx@71\msx@71 }
\def\llcorner{\delimiter"4\msx@78\msx@78 }
\def\lrcorner{\delimiter"5\msx@79\msx@79 }
\def\yen{\mathhexbox\msx@55 }
\def\checkmark{\mathhexbox\msx@58 }
\def\circledR{\mathhexbox\msx@72 }
\def\maltese{\mathhexbox\msx@7A }
\mathchardef\lvertneqq="3\msy@00
\mathchardef\gvertneqq="3\msy@01
\mathchardef\nleq="3\msy@02
\mathchardef\ngeq="3\msy@03
\mathchardef\nless="3\msy@04
\mathchardef\ngtr="3\msy@05
\mathchardef\nprec="3\msy@06
\mathchardef\nsucc="3\msy@07
\mathchardef\lneqq="3\msy@08
\mathchardef\gneqq="3\msy@09
\mathchardef\nleqslant="3\msy@0A
\mathchardef\ngeqslant="3\msy@0B
\mathchardef\lneq="3\msy@0C
\mathchardef\gneq="3\msy@0D
\mathchardef\npreceq="3\msy@0E
\mathchardef\nsucceq="3\msy@0F
\mathchardef\precnsim="3\msy@10
\mathchardef\succnsim="3\msy@11
\mathchardef\lnsim="3\msy@12
\mathchardef\gnsim="3\msy@13
\mathchardef\nleqq="3\msy@14
\mathchardef\ngeqq="3\msy@15
\mathchardef\precneqq="3\msy@16
\mathchardef\succneqq="3\msy@17
\mathchardef\precnapprox="3\msy@18
\mathchardef\succnapprox="3\msy@19
\mathchardef\lnapprox="3\msy@1A
\mathchardef\gnapprox="3\msy@1B
\mathchardef\nsim="3\msy@1C
\mathchardef\napprox="3\msy@1D
\mathchardef\nsubseteqq="3\msy@22
\mathchardef\nsupseteqq="3\msy@23
\mathchardef\subsetneqq="3\msy@24
\mathchardef\supsetneqq="3\msy@25
\mathchardef\subsetneq="3\msy@28
\mathchardef\supsetneq="3\msy@29
\mathchardef\nsubseteq="3\msy@2A
\mathchardef\nsupseteq="3\msy@2B
\mathchardef\nparallel="3\msy@2C
\mathchardef\nmid="3\msy@2D
\mathchardef\nshortmid="3\msy@2E
\mathchardef\nshortparallel="3\msy@2F
\mathchardef\nvdash="3\msy@30
\mathchardef\nVdash="3\msy@31
\mathchardef\nvDash="3\msy@32
\mathchardef\nVDash="3\msy@33
\mathchardef\ntrianglerighteq="3\msy@34
\mathchardef\ntrianglelefteq="3\msy@35
\mathchardef\ntriangleleft="3\msy@36
\mathchardef\ntriangleright="3\msy@37
\mathchardef\nleftarrow="3\msy@38
\mathchardef\nrightarrow="3\msy@39
\mathchardef\nLeftarrow="3\msy@3A
\mathchardef\nRightarrow="3\msy@3B
\mathchardef\nLeftrightarrow="3\msy@3C
\mathchardef\nleftrightarrow="3\msy@3D
\mathchardef\divideontimes="2\msy@3E
\mathchardef\varnothing="0\msy@3F
\mathchardef\nexists="0\msy@40
\mathchardef\mho="0\msy@66
\mathchardef\thorn="0\msy@67
\mathchardef\beth="0\msy@69
\mathchardef\gimel="0\msy@6A
\mathchardef\daleth="0\msy@6B
\mathchardef\lessdot="3\msy@6C
\mathchardef\gtrdot="3\msy@6D
\mathchardef\ltimes="2\msy@6E
\mathchardef\rtimes="2\msy@6F
\mathchardef\shortmid="3\msy@70
\mathchardef\shortparallel="3\msy@71
\mathchardef\smallsetminus="2\msy@72
\mathchardef\thicksim="3\msy@73
\mathchardef\thickapprox="3\msy@74
\mathchardef\approxeq="3\msy@75
\mathchardef\succapprox="3\msy@76
\mathchardef\precapprox="3\msy@77
\mathchardef\curvearrowleft="3\msy@78
\mathchardef\curvearrowright="3\msy@79
\mathchardef\digamma="0\msy@7A
\mathchardef\varkappa="0\msy@7B
\mathchardef\hslash="0\msy@7D
\mathchardef\hbar="0\msy@7E
\mathchardef\backepsilon="3\msy@7F
\def\Bbb{\ifmmode\let\next\Bbb@\else
 \def\next{\errmessage{Use \string\Bbb\space only in math mode}}\fi\next}
\def\Bbb@#1{{\Bbb@@{#1}}}
\def\Bbb@@#1{\fam\msyfam#1}

\def\inv{^{\raise.15ex\hbox{${
  \scriptscriptstyle -}$}\kern-.05em 1}}

\def\Dsl{\,\raise.15ex\hbox{$/$}\mkern-13.5mu D}
\def\dsl{\raise.15ex\hbox{$/$}\kern-.57em\hbox{$\partial$}}

\def\lspace{\ifx\answ\bigans{}\else\qquad\fi}
\def\del{\partial}
 \def\CC{\hbox{{$\cal C$}}}
 
\def\CL{\hbox{{$\cal L$}}} 
 
 \def\CS{\hbox{{$\cal S$}}}

\def\CR{\hbox{{$\cal R$}}} 
\def\CT{\hbox{{$\cal T$}}}

\def\CO{\hbox{{$\cal O$}}}

\def\lform{\hbox{$\sqcup$}\llap{\hbox{$\sqcap$}}}
\def\darr#1{\raise1.5ex\hbox{$\leftrightarrow$}
\mkern-16.5mu #1}

\def\h{{{1\over2}}}

\def\INT{{\textstyle \int\kern-.642em\int}}

\def\R{{\Bbb R}}
\def\C{{\Bbb C}}
\def\Z{{\Bbb Z}}

\def\eps{{\epsilon}}

\def\aut{{\rm Aut\, }}

\def\ant{{{\scriptstyle S}}}

\def\cocross{{>\!\!\!\triangleleft}}

\def\tens{\mathop{\otimes}}

\def\la{{\triangleright}}\def\ra{{\triangleleft}}

\def\isom{{\cong}}

\def\sgn{{\rm sgn}}

\def\Hom{{\rm Hom}}

\def\Ad{{\rm Ad}}

\def\sgn{{\rm sgn}}

\def\id{{\rm id}}
\def\Deltaop{{\Delta^{\rm op}}}

\def\nquad{{\!\!\!\!\!\!}}
\def\nqquad{\nquad\nquad}
\def\eqn#1#2{\begin{equation}#2\label{#1}\end{equation}}

\def\o{{}_{(1)}}\def\t{{}_{(2)}}\def\th{{}_{(3)}}

\def\bo{{}^{\bar{(1)}}}\def\bt{{}^{\bar{(2)}}}

\def\und#1{{\underline {#1}}}

\def\uo{{{}^{(1)}}}\def\ut{{{}^{(2)}}}

\def\uth{{{}^{(3)}}}

\def\new#1{\goodbreak\goodbreak\bigskip
\noindent{\bf #1}}

\def\text#1{\mbox{\rm #1}}
\def\note#1{}

\def\blacksquare{{\lform}}
\def\frac#1#2{{{#1\over#2}}}

\def\proof{\goodbreak\noindent{\bf Proof\quad}}
\def\endproof{{\ $\lform$}\bigskip }

\def\align#1{\begin{eqnarray*}#1\end{eqnarray*}}

\def\alignn#1#2{\begin{eqnarray}\label{#1}#2
\end{eqnarray}}

\def\und#1{{\underline{#1}}}

\def\usl{{U_q(sl_2)}}
\def\busl{{B{\usl}}}
\def\so{{{}_{[1]}}}\def\st{{{}_{[2]}}}

\def\<{\langle}
\def\>{\rangle}

\makeatother

\documentstyle[11pt]{article}

\def\thebibliography#1{\section*{REFERENCES}\list
 {[\arabic{enumi}]}{\settowidth\labelwidth{[#1]}\leftmargin\labelwidth
 \advance\leftmargin\labelsep
 \usecounter{enumi}}
 \def\newblock{\hskip .11em plus .33em minus -.07em}
 \sloppy
 \sfcode`\.=1000\relax}

\newtheorem{lemma}{Lemma}[section]
\newtheorem{propos}[lemma]{Proposition}

\newtheorem{theorem}[lemma]{Theorem}

\newtheorem{corol}[lemma]{Corollary}
\newtheorem{defin}[lemma]{Definition}

\begin{document}\baselineskip 20pt

{\ }\hskip 4.7in DAMTP/92-10-revised
\vspace{0in}

\begin{center} {\bf BRAIDED MATRIX STRUCTURE OF THE SKLYANIN ALGEBRA AND OF THE
QUANTUM LORENTZ GROUP}
\baselineskip 13pt{\ }\\
{\ }\\ S. Majid\footnote{SERC Fellow and Drapers Fellow of Pembroke College,
Cambridge}\\ {\ }\\
Department of Applied Mathematics\\
\& Theoretical Physics\\ University of Cambridge\\ Cambridge CB3 9EW, U.K.
\end{center}

\begin{center}
January 1992 \end{center}
\vspace{10pt}
\begin{quote}\baselineskip 13pt
\noindent{\bf ABSTRACT} Braided groups and braided matrices are novel algebraic
structures living in braided or quasitensor categories. As such they are a
generalization of super-groups and super-matrices to the case of braid
statistics. Here we construct braided group versions of the standard quantum
groups $U_q(g)$. They have the same FRT generators $l^\pm$ but a matrix
braided-coproduct $\und\Delta L=L\und\tens L$ where $L=l^+Sl^-$, and are
self-dual. As an application, the degenerate Sklyanin algebra is shown to be
isomorphic to the braided matrices $BM_q(2)$; it is a braided-commutative
bialgebra in a braided category. As a second application, we show that the
quantum double $D(\usl)$ (also known as the `quantum Lorentz group') is the
semidirect product as an algebra of two copies of $\usl$, and also a semidirect
product as a coalgebra if we use braid statistics. We find various results of
this type for the doubles of general quantum groups and their semi-classical
limits as doubles of the Lie algebras of Poisson Lie groups.
\end{quote}
\baselineskip 22pt

\section{INTRODUCTION}

Historically, the existence of particles with bose and fermi statistics led
physicists naturally to the study of super-algebras and super-groups. In a
similar way, the existence in low-dimensional quantum field theory of particles
with braid statistics\cite{FRS:sup}\cite{Lon:ind} surely motivates the study of
novel braided algebraic structures. The formulation and study of precisely such
new algebraic structures has been initiated in
\cite{Ma:som}\cite{Ma:bra}\cite{Ma:rec}\cite{Ma:exa}\cite{Ma:eul}\cite{Ma:sta}
\cite{Ma:any}\cite{Ma:csta} and
\cite{Ma:bg}\cite{Ma:tra}\cite{Ma:bos}\cite{Ma:cat}
under the heading `braided groups'. They precisely generalize results
about super-algebras and super-groups to a situation in which the
super-transposition map $\Psi(b\tens c)=(-1)^{|b||c|}c\tens b$ on
homogeneous elements, is replaced by a braided-transposition or
braiding $\Psi$ obeying the Yang-Baxter equations. This is formulated
mathematically by means of the theory of braided or quasitensor
categories and it is in such a category that a braided group lives
(just as a super-algebra or super-Lie algebra lives in the category of
super-spaces). Among the general results is that in every situation
where there is a quasitensor category $\CC$ (such as the
super-selection structure in low-dimension quantum theory) there is to
be found a braided group $\aut(\CC)$\cite{Ma:bg}. So such objects are
surely relevant to physics\cite{Ma:bra}\cite{Ma:sta}.

The motivation for this theory so far is however, of a very general nature.
Here we develop two very specific applications of the general theory to the
structure of two algebras of independent interest in physics, namely to the
degenerate Sklyanin algebra and to the quantum Lorentz group, or more generally
to the quantum double of a general quantum group. We will see that these
algebraic structures are naturally endowed with braid statistics and that this
braiding enables us to obtain new results about them. We also obtain in the
semiclassical limit a new result about Poisson Lie algebras.

The precise definition of a braided group is recalled in the Preliminaries
below. The first step is to formulate precisely what we mean by braid
statistics using the theory of quasitensor categories. This is a collection of
objects closed under a tensor product $\tens$ and equipped with isomorphisms
$\Psi_{V,W}:V\tens W\to W\tens V$ for any two objects $V$ and $W$. These play
the role of the usual transposition for vector spaces or the
super-transposition for super-spaces mentioned above. The second idea is to
remember that instead of
working with groups or Lie algebras we can work with their corresponding
cocommutative group or enveloping Hopf algebras. The same is known in the
super-case where we can work with super-cocommutative Hopf algebras rather than
with the super-group or super-Lie algebra itself. Likewise in the braided case
we work directly with braided-cocommutative Hopf algebras living in a
quasitensor category. We call such objects braided groups of enveloping algebra
type. So far, there does not appear to be any general theory of braided Lie
algebras themselves underlying these braided groups (we make some remarks about
this in Section~2).

Because we are working with Hopf algebras (albeit living in a quasitensor
category), the mathematical technology here is for the most part already
familiar to physicists in the context of quantum groups. Thus, we have an
algebra $B$ and a braided-coproduct as a coassociative homomorphism
$\und\Delta:B\to B\und \tens B$. The crucial difference is that $B\und\tens B$
is not an ordinary tensor product but a braided one. It coincides with $B\tens
B$ as an object but its two subalgebras $B$ do not commute, enjoying instead an
exchange rule given by $\Psi$. Thus
\[ (a\tens b)(c\tens d)=a\Psi(b\tens c)d\]
where $\Psi_{B,B}:B\tens B\to B\tens B$ is a Yang-Baxter operator. This is
clearly a generalization of the super-tensor product of super-algebras and
shows the use of $\Psi$ as a kind of braid statistics. The braided tensor
product construction here is very natural from the point of view of algebras
living in a quasitensor category. Let us note that such quasitensor categories
are closely related to the theory of link
invariants\cite{FreYet:bra}\cite{ResTur:rib} and indeed many of the abstract
braided group computations and proofs are best done by means of drawing braids
and tangles, see \cite{Ma:sta}\cite{Ma:tra}\cite{Ma:bos}. For example, the
proof of associativity of $B\und\tens B$ (or more generally $B\und\tens C$ for
two algebras in the category) depends on the functoriality and hexagon
identities for $\Psi$ and is most easily done by such diagrammatic means. This
is one of the novel features of these new algebraic structures.

Among the general results in \cite{Ma:bra}\cite{Ma:tra} is that every quantum
group $(H,\CR)$ (by this we mean an ordinary Hopf algebra equipped with a
quasitriangular structure or `universal $R$-matrix'\cite{Dri}, such as the
$U_q(g)$ of Drinfeld and Jimbo\cite{Dri}\cite{Jim:dif}) can be transmuted in a
canonical way into a braided group. Only the coalgebra need be changed and
$\Psi$ is obtained in a standard way from $\CR$ in the adjoint representation.
Likewise every dual quantum group $(A,\CR)$ (by which we mean a
dual-quasitriangular Hopf lagebra such as the quantum function algebra
$SL_q(2)$) can be transmuted to a braided-group of function algebra type. This
is braided-commutative in a certain sense (rather than braided-cocommutative as
above). By means of these transmutation processes we can obviously formulate
all questions about quantum groups in terms of their associated braided groups.
In brief, braided groups generalize both super groups and (via transmutation)
quantum groups.

We begin in Section~2 by computing the braided groups $BU_q(g)$ of enveloping
algebra type associated to the familiar quantum groups $U_q(g)$. They turn out
to have the same generators $l^\pm$ of $U_q(g)$ in FRT form\cite{FRT:lie} but a
new braided-coproduct. Writing $L=l^+\ant l^-$, the braided-coproduct comes out
as the matrix one $\und\Delta L=L\und\tens L$. Here $\ant$ is the antipode for
the quantum group. Exactly this combination of generators $L=l^+\ant l^-$ is
well-known in certain contexts\cite{ResSem:cen}\cite{MorRes:com}\cite{AFS:hid}
but until now the obvious matrix coalgebra $\und\Delta L=L\und\tens L$ has had
no role (the usual coproduct of $U_q(g)$ is not so simple in terms of $L$). We
see now that in the braided setting these generators are very natural. Our
matrix braided-coproduct can be used in all the same ways as the usual
coproduct provided only that one remembers always to work in the quasitensor
category (by using the relevant $\Psi$ in place of any usual transpositions).

Essential in this computation is the fact that these quantum groups $U_q(g)$
are factorizable in the sense\cite{ResSem:mat} that they have a non-degenerate
`quantum Killing form' $Q=\CR_{21}\CR_{12}$. We also note, perhaps
surprisingly, that the corresponding braided groups are self-dual: the
$BU_q(g)$ are isomorphic via $Q$ to the corresponding braided groups of
function algebra type already computed in \cite{Ma:exa} from an $R$-matrix.
This is a purely quantum phenomenon in that it holds only for generic $q\ne 1$
and certain roots of unity. It has important consequences, some of which are
already known in other contexts. For example, it means that the braided
versions of the familiar quantum groups are both braided-cocommutative and
braided-commutative and indeed self-dual, so more like $\R^n$ than anything
else\cite{LyuMa:bra}\cite{LyuMa:fou}. Related to this, though we do not
describe an abstract notion of braided Lie algebras, we note that the
generators $L$ of $BU_q(g)$ do enjoy a kind of Lie bracket (which we compute)
based on the quantum adjoint action and obeying some Lie-algebra like
identities.

Our first main application is in Section~3 where we study the degenerate
Sklyanin algebra. The Sklyanin algebra was introduced in \cite{Skl:alg} as a
way to generate representations of a certain bialgebra (by which we mean a Hopf
algebra without antipode) related to the 8-vertex model. Apart from that, it
has remarkable mathematical properties (for example, the same Poincar\'e series
as the commutative ring of polynomials in 4 variables\cite{SmiSta:reg}) and is
related to the theory of elliptic curves. It has three parameters governed by
one constraint. The degenerate case in which one of the parameters vanishes has
been of fundamental importance in the development of quantum groups because a
quotient of it led to the quantum group $U_q(sl_2)$. Many authors have wondered
accordingly if the Skylanin algebra is also a quantum group or bialgebra. So
far this has defied all attempts even in the degenerate case: the Skylanin
algebra does not appear to be any kind of usual quantum group or bialgebra. We
show that the degenerate Sklyanin algebra is, however, a bialgebra living in a
quasitensor category. It is in fact isomorphic to the braided matrices
$BM_q(2)$ introduced in \cite{Ma:exa}\cite{Ma:eul} with braided-coproduct again
in matrix from. This means that it is indeed some kind of group-like object in
the sense that we can tensor product its (braided) representations, act by it
on algebras etc, just as we can for any group or quantum group. Moreover, the
quotienting procedure to obtain the algebra of $U_q(sl_2)$ from the degenerate
Sklyanin algebra is understood now as setting equal to one the braided
determinant of the braided-matrix generator.

Our second main application is in Section~4 where we study the quantum double
of a quantum group, for example the quantum double of $U_q(sl_2)$. The quantum
double $D(H)$ is a general construction for obtaining a new quantum group from
any Hopf algebra $H$\cite{Dri}. It is built on the linear space $H\tens H^*$
with a doubly-twisted product (here $H^*$ is dual to $H$). The case when $H$ is
quasitriangular was studied explicitly by the author in \cite{Ma:dou} and we
shall build on the results there concerning the semidirect product structure of
$D(H)$ in this case. The quantum double in this case, particularly $D(U_q(g))$
is of considerable interest in physics. One of the reasons for interest, which
we will be able to illuminate, is the idea in \cite{PodWor:def} that such a
double should be regarded as a kind of complexification of the quantum group
$U_q(g)$. For, example $D(U_q(sl_2))$ should be regarded as (by definition) the
quantum enveloping algebra of the Lorentz group $so(1,3)$. \cite{PodWor:def}
obtained some arguments for this in the dual $C^*$-algebra context of matrix
pseudogroups, but some puzzles remain regarding such an interpretation. In
particular, the quantum double is {\em a priori} built on the tensor product of
a quantum group of enveloping algebra type and one of function algebra type
(dual to the first). So $D(U_q(sl_2))$ does not look at first like two copies
of $U_q(sl_2)$. By means of results in Section~2 and the theory of braided
groups we show that in fact $D(U_q(g))$  is indeed generated as an algebra by
two copies of $U_q(g)$, with certain cross relations: it is a semidirect
product of one copy acting on the other by the quantum adjoint action
\[ D(U_q(g))\isom U_q(g){}_{\Ad}\cocross U_q(g)\]
as an algebra. Explicitly, if $l^\pm$ are the FRT generators of one copy and,
say $m^\pm$ of the other then the cross relations are
\[ R_{12}l_2^+M_1=M_1 R_{12}l_2^+,\quad R_{21}^{-1}l_2^-
M_1=M_1R_{21}^{-1}l_2^-\]
where $M=m^+\ant m^-$. The notation here is the standard one for matrix
generators and $R$ is the appropriate $R$-matrix for the quantum group. This
gives a very simple matrix description of the algebra of $D(U_q(g))$. Moreover,
we show that the coalgebra of the quantum double in our description also has a
semidirect coproduct form, namely
\[ \Delta l^\pm=l^\pm\tens l^\pm,\quad \Delta M=(\sum M\CR\ut\tens \CR\uo
M)\CR^{-1}_{21}.\]
Here $\CR=\sum\CR\uo\tens\CR\ut$ lives in the tensor square of the copy of
$U_q(g)$ generated by $l^\pm$. Note here the use of the braided-coproduct
$\und\Delta M=M\und\tens M$. These results about the structure of the quantum
double are obtained in Corollary~4.3 as the assertion that
\[ D(U_q(g))\isom BU_q(g)\cocross U_q(g)\]
as an algebra and coalgebra. This also implies that the dual of the quantum
double is also isomorphic to a semidirect product, a fact which is far from
evident in \cite{PodWor:def}. In general, such a semidirect product structure
for the quantum double has far reaching consequences. For example its
representation theory can be analysed by a Hopf algebra version of Mackey's
construction for semidirect products. It is also relevant to recent approaches
to the quantum differential calculus on quantum groups. Let us note that the
quantum double has also been studied in \cite{FRT:dou}\cite{ResSem:mat} among
other places.

Here we concentrate on illuminating the semidirect product result by computing
its semiclassical version. This is also to be found in Section~4. The
semiclassical notion of a quantum group is a quasitriangular Lie bialgebra
$(g,r)$ as introduced by Drinfeld\cite{Dri:ham}. Here $g$ is a Lie algebra and
$r\in g\tens g$ obeys the Classical Yang-Baxter equations. The corresponding
Lie group has a compatible Poisson bracket obtained from $r$ (it is a Poisson
Lie group). In some sense, $U_q(g)$ is a `quantization' of such a $(g,r)$.
There is a classical double construction $D(g)$ also introduced by
Drinfeld\cite{Dri:ham} built on $g\oplus g^*$ with  a doubly-twisted Lie
bracket. For the standard Drinfeld-Jimbo $r$-matrix it is known that $g^*$ is a
solvable Lie algebra (hence quite different from $g$) while $D(g)$ has a real
form isomorphic to the complexification $g_\C$ and $g,g^*$ are components in
its Iwasawa decomposition. Factorizability of $U_q(g)$ corresponds at the
semiclassical level to the linear isomorphism $K:g^*\to g$ provided by the
Killing form when $g$ is semisimple. Our semidirect product theorem for the
quantum double now becomes
\[ D(g)\isom g\cocross g\]
whenever $g$ is a quasitriangular Lie bialgebra with non-degenerate symmetric
part of $r$.  This result is obtained in Theorem~4.4 and is related to the
Iwasawa decomposition of $g_\C$ in Corollary~4.5. Thus, at least at the
semiclassical level our semidirect product result is compatible with the view
of $D(g)$ as complexification of $g$.

This completes our outline of the main results of the paper. For completeness
we also include some related results about the quantum double in relation to
braided statistics. Firstly, one can replace the second $U_q(g)$ also by its
braided version: we show that the semidirect product $BU_q(g)\cocross BU_q(g)$
is an example of a quantum braided group (i.e. a quasitriangular Hopf algebra
living a quasitensor category). As a coalgebra it is now a braided-tensor
coproduct as explained in Corollary~4.7. Finally, because $D(H)$ is a quantum
group, it too has its own associated braided group $B\! D(H)$. We compute this
in the Appendix, with emphasis on the simplest case where $H=\C G$ the group
algebra of a finite group. The importance of this structure in physics is less
well-established but it can be expected to play a role in Chern-Simons theories
with finite gauge group as in \cite{DijWit:top} or in theories exhibiting
non-Abelian anyon statistics. We also give in a braided interpretation of an
old theorem of Radford\cite{Rad:str} to the effect that if
$H_1{\to\atop\hookleftarrow} H$ is any Hopf algebra projection then $H_1\isom
B\cocross H$ where $B$ is a Hopf algebra living (like $B\! D(H)$ above) in the
quasitensor category of $D(H)$-representations. This kind of Hopf algebra
projection arises for certain quantum homogeneous spaces.

\new{PRELIMINARIES}

Braided monoidal or quasitensor categories have been formally introduced into
category theory in \cite{JoyStr:bra}. A quasitensor category means for us
$(\CC,\tens,\und1,\Phi,\Psi)$ where $\CC$ is a category equipped with a
monoidal product $\tens$ and identity object $\und 1$ (with some associated
maps) and functorial associativity isomorphisms $\Phi_{V,W,Z}:V\tens (W\tens
Z)\to (V\tens W)\tens Z$ for any three objects $V,W,Z$, obeying Maclane's
pentagon identity (so ($\CC,\tens,\und1,\Phi)$ is a monoidal category). In
addition for a quasitensor category we need functorial quasi-symmetry
isomorphisms $\Psi_{V,W}:V\tens W\to W\tens V$ obeying two hexagon identities.
If we omit $\Phi$ (it will be trivial in the examples of interest here) then
the hexagons take the form
\eqn{hex}{ \Psi_{V,W\tens Z}=\Psi_{V,Z}\circ\Psi_{V,W},\quad\Psi_{V\tens
W,Z}=\Psi_{V,Z}\circ\Psi_{W,Z}.}
One can deduce that $\Psi_{V,\und 1}=\id=\Psi_{\und 1,V}$ for any $V$. If
$\Psi_{V,W}=\Psi^{-1}_{W,V}$ for all $V,W$ ( $\Psi^2=\id$) then one of the
hexagons is superfluous and
we have an ordinary symmetric monoidal or tensor
category as in \cite{Mac:cat}. In general $\Psi_{V,W},\Psi^{-1}_{W,V}$ are
distinct and commonly represented as distinct  braid crossings connecting
$V\tens W$ to $W\tens V$. The coherence theorem for quasitensor categories says
that if a sequence of the $\Psi,\Psi^{-1}$ written in this way correspond to
the same braid then they compose to the same map. The quasisymmetry $\Psi$ can
be called a `braiding' or `braided-transposition' for this reason.
Functoriality of $\Psi$ asserts explicitly that $\Psi$ commutes with morphisms
in the sense
\eqn{funct}{\Psi_{V,Z}\circ(\id\tens \phi)=(\phi\tens\id)\circ\Psi_{V,W},\quad
\forall \phi:W\to Z}
on one input and similarly on the other input. In the diagrammatic notation,
functoriality for $\Psi,\Psi^{-1}$ asserts that any morphisms between objects
commute in the sense that they can be pulled through braid crossings. Here a
typical morphism with $n$ inputs and $m$ outputs is represented as an
$n+m$-vertex. We draw all morphisms pointing downwards.

The idea of an algebra in a quasitensor category (indeed, in any monoidal one)
is just the obvious one: it is an object $B$ in the category and morphisms
$\und\eta:\und 1\to B$, $\und\cdot:B\tens B\to B$ obeying the usual axioms (as
diagrams) for
associativity and unity. We use the term `algebra' a little loosely here,
however in our examples, all constructions will indeed be $k$-linear over a
field or ring $k$. Of crucial importance for the theory of braided groups is
that if $B,C$ are two algebras in a quasitensor category then one can define a
new algebra, which we call the {\em braided tensor product algebra} $B\und\tens
C$ also living in the category. As an object it consists of $B\tens C$ equipped
now with the algebra structure
\eqn{bratensprod}{\und\cdot_{B\und\tens
C}=(\und\cdot_{B}\tens\und\cdot_{C})(\id\tens\Psi_{C,B}\tens\id).}
The proof that this is associative is a good exercise in the diagrammatic
notation mentioned above\cite{Ma:sta}. For completeness, we have recalled the
relevant diagram in Figure~1. The first equality is functoriality under the
morphism $\und\cdot:B\tens B\to B$, the second is associativity of the products
in $B,C$ and the third is functoriality under $\und\cdot:C\tens C\to C$. There
is another opposite braided tensor product using $\Psi^{-1}_{B,C}$ instead of
$\Psi_{C,B}$.

\begin{figure}
\unitlength=1.00mm
\special{em:linewidth 0.4pt}
\linethickness{0.4pt}
\begin{picture}(100.00,111)(0,28)
\put(30.00,134.00){\makebox(0,0)[cc]{$(B\tens C)\tens (B\tens C)\tens (B\tens
C)$}}
\put(90.00,134.00){\makebox(0,0)[cc]{$(B\tens C)\tens (B\tens C)\tens (B\tens
C)$}}
\put(40.00,73.00){\makebox(0,0)[cc]{$(B\tens C)\tens (B\tens C)\tens (B\tens
C)$}}
\put(100.00,73.00){\makebox(0,0)[cc]{$(B\tens C)\tens (B\tens C)\tens (B\tens
C)$}}
\put(30.00,89.00){\makebox(0,0)[cc]{$(B\tens C)$}}
\put(90.00,89.00){\makebox(0,0)[cc]{$(B\tens C)$}}
\put(40.00,28.00){\makebox(0,0)[cc]{$(B\tens C)$}}
\put(100.00,28.00){\makebox(0,0)[cc]{$(B\tens C)$}}
\put(60.00,110.00){\makebox(0,0)[cc]{$=$}}
\put(70.00,49.00){\makebox(0,0)[cc]{$=$}}
\put(10.00,49.00){\makebox(0,0)[cc]{$=$}}
\end{picture}
\caption{Proof of associativity of the braided tensor product of algebras}
\end{figure}
\begin{figure}
\unitlength=1.00mm
\special{em:linewidth 0.4pt}
\linethickness{0.4pt}
\begin{picture}(129.00,40.00)(0,110)
\put(15.00,140.00){\makebox(0,0)[cc]{$B\tens B$}}
\put(15.00,110.00){\makebox(0,0)[cc]{$B\tens B$}}
\put(50.00,140.00){\makebox(0,0)[cc]{$B\tens B$}}
\put(50.00,110.00){\makebox(0,0)[cc]{$B\tens B$}}
\put(93.00,140.00){\makebox(0,0)[cc]{$B\tens V$}}
\put(93.00,110.00){\makebox(0,0)[cc]{$B\tens V$}}
\put(129.00,140.00){\makebox(0,0)[cc]{$B\tens V$}}
\put(129.00,110.00){\makebox(0,0)[cc]{$B\tens V$}}
\put(30.00,125.00){\makebox(0,0)[cc]{$=$}}
\put(113.00,125.00){\makebox(0,0)[cc]{$=$}}
\end{picture}
\caption{(a) Hopf algebra axiom. (b) Opposite coproduct axiom}
\end{figure}

This braided tensor product is the crucial ingredient in the definition of a
Hopf algebra living in a quasitensor
category\cite{Ma:som}\cite{Ma:bra}\cite{Ma:rec}\cite{Ma:exa}\cite{Ma:eul}
\cite{Ma:sta}. This means $(B,\und\Delta,\und\eps,\und\ant)$ where $B$ is an
algebra living in the category and $\und\Delta:B\to B\und\tens B$,
$\und\eps:B\to\und 1$ are algebra homomorphisms. In addition
$(B,\und\Delta,\und\eps)$ form a coalgebra in the usual way (so
$(\und\Delta\tens\id)\und\Delta=(\id\tens\und\Delta)\und\Delta$,
$(\und\eps\tens\id)\und\Delta=\id=(\id\tens\und\eps)\und\Delta$).
The antipode $\und \ant$ if it exists also obeys the usual axioms (so
$\und\cdot(\und\ant\tens\id)\und\Delta=\und\eta\und\eps=\und\cdot
(\id\tens\und\ant)\und\Delta$). These axioms are analogous to the usual axioms
for an Hopf algebra as in \cite{Swe:hop} but now as morphisms in the category.
If there is no antipode then we have merely a bialgebra in the category. The
diagrammatic form for the bialgebra axiom for $\und\Delta$ is shown in
Figure~2(a).

As explained in the introduction, a {\em braided group} means a Hopf algebra
living in a quasitensor category, equipped with some further structure
expressing some kind of braided-commutativity or braided-cocommutativity (so
that it is like the function algebra or enveloping algebra respectively of a
classical group). The notion of braided (co)-commutativity that we need is a
little involved in the abstract setting (though it is clear enough in the
concrete examples that we need below). The problem in the abstract setting is
that the obvious notions of opposite product
$\und\cdot\circ\Psi_{B,B},\und\cdot\circ\Psi^{-1}_{B,B}$ or opposite coproduct
$\Psi_{B,B}\circ \und\Delta,\Psi^{-1}_{B,B}\circ\und\Delta$ do not again define
Hopf algebras in the quasitensor category when $\Psi^2\ne\id$. For example,
$\Psi_{B,B}^{-1}\circ \und\Delta:B\to B\tens B$ gives a homomorphism to the
opposite braided tensor product algebra, and hence a Hopf algebra in the
quasitensor category with opposite braiding. Again, this is easy to see using
the diagrammatic notation. Since there is no intrinsic notion of opposite Hopf
algebra in the braided case, there is no intrinsic way to assert that a Hopf
algebra in the category coincides with its opposite.

In practice, we avoid this problem by working with a weaker (non-intrinsic)
notion of braided-commutativity or braided-cocommutativity defined with respect
to a class $\CO$ of comodules or modules respectively. Here a $B$-module in the
category is $(V,\alpha_V)$ where $V$ is an object and $\alpha_V:B\tens V\to V$
is a morphism such that $\alpha_V(\id\tens
\alpha_V)=\alpha_V(\und\cdot\tens\id)$, $\alpha_V(\und\eta\tens\id)=\id$ as
usual. Then

\begin{defin}\cite{Ma:bra}\cite{Ma:tra} A (weak) opposite coproduct for a
bialgebra $B$ in a quasitensor category, is a pair $(\und\Deltaop,\CO)$ where
$\und\Deltaop:B\to B\und\tens B$ defines a second bialgebra structure for the
same algebra $B$, and $\CO$ is a class of $B$-modules such that the condition
in Figure~2(b) holds for all $(V,\alpha_V)$ in the class.
\end{defin}

\begin{defin}\cite{Ma:bra}\cite{Ma:tra} A braided-cocommutative bialgebra is a
pair $(B,\CO)$ where $B$ is a bialgebra in the category and $(\und\Delta,\CO)$
is an opposite coproduct. A braided group (of enveloping algebra type) is a
pair $(B,\CO)$ where $B$ is a Hopf algebra in a quasitensor category and
$(B,\CO)$ is braided-cocommutative.\end{defin}

Thus a braided group is nothing other than a Hopf algebra $B$ in a quasitensor
category equipped with a class $\CO$ with respect to which it behaves in a
cocommutative way. In our examples, the class $\CO$ is quite large and contains
all useful modules (such as the braided-adjoint action of any Hopf algebra in
the category on itself). For a given Hopf algebra $B$ the class of all modules
$\CO(B)$ with respect to which it is cocommutative is closed under tensor
product\cite[Theorem~3.2]{Ma:tra} and under dualization if the quasitensor
category has duals \cite[Theorem~3.1]{Ma:tra}. The notion of opposite product
is analogous, working now with a class of comodules in the category obeying an
analogous condition (its diagram is just given by turning Figure~2(b) upside
down). It enables us to similarly define a braided-group of function algebra
type as a braided-commutative Hopf algebra in a quasitensor category.

The above constructions in quasitensor categories are necessarily quite
abstract. Fortunately, we will be concerned below only with the quasitensor
categories that arise as representations of a quantum group, in which case
explicit formulae are possible. Here a quantum group means for us the data
$(H,\Delta,\eps,\ant,\CR)$ where $\Delta:H\to H\tens H$ is the coproduct,
$\eps:H\to k$ the counit and $\ant:H\to H$ the antipode forming an ordinary
Hopf algebra over a field (or, with care, a commutative ring) $k$. For these we
use the standard notation of \cite{Swe:hop},
notably $\Delta h=\sum h\o\tens h\t$ for the action of $\Delta$ on $h\in H$.
The additional invertible element
$\CR\in H\tens H$ is the quasitriangular structure or `universal R-matrix' and
obeys the axioms of Drinfeld\cite{Dri}
\eqn{qtri}{ (\Delta\tens\id)\CR=\CR_{13}\CR_{23},\qquad
(\id\tens\Delta)\CR=\CR_{13}\CR_{12},\quad \Deltaop=\CR(\Delta\ )\CR^{-1}}
where $\Deltaop$ denotes $\Delta$ followed by the usual transposition on
$H\tens H$, and $\CR_{12}=\CR\tens 1$ in $H^{\tens 3}$ as usual. A quantum
group is said to be {\em triangular} if $\CR_{21}\CR_{12}=1\tens 1$\cite{Dri}
and strictly quasitriangular otherwise. We refer to \cite{Dri}\cite{Jim:dif}
for the standard (strictly) quasitriangular Hopf algebras $U_q(g)$. At generic
$q=e^{\hbar\over 2}$ these are viewed over the ring of formal power series
$\C[[\hbar]]$\cite{Dri}.

These axioms are indeed such as to ensure that the category of representations
of a quantum group form a quasitensor category. Some treatments of this topic
appeared independently in \cite[Sec. 7]{Ma:qua} and \cite{ResTur:rib} as well
as surely being known to experts at the time.
Briefly, the objects in this category are the representations (modules) of $H$,
while the tensor product of two such representations is given in the usual way
by pull back along $\Delta$. For any two such objects $V,W$ the quasisymmetry
or braiding is given by
\eqn{e1}{\Psi_{V,W}(v\tens w)=\sum \CR\ut\la w\tens \CR\uo\la v}
where $\CR=\sum\CR\uo\tens \CR\ut$ and $\la$ denotes the relevant actions. One
can easily check that $\Psi_{V,W}:V\tens W\to W\tens V$ obeys (\ref{hex}) and
is indeed an intertwiner for the action of $H$, precisely because $\CR$ obeys
(\ref{qtri}). Moreover, the category is an ordinary symmetric one (with
$\Psi^2=\id$) precisely when the quantum group is triangular rather than
strictly triangular. The role of the quantum group here is to generate the
quasitensor category in which we work. For example, the finite-dimensional
quantum group $(\Z_2',\CR)$ in \cite[Sec. 6]{Ma:exa} generates the category of
super-vector spaces as its representations. Here $\Z_2'$ denotes the group
algebra of $\Z_2$ equipped with a certain non-trivial $\CR$.

In \cite{Ma:bra}\cite{Ma:eul}\cite{Ma:bg} we gave a construction for braided
groups in this kind of quasitensor category. In fact, there is a canonical
construction for each type beginning from the quantum group itself
that generates the category. Thus, if $H$ is a quantum group, it has an
associated braided group (of enveloping algebra type) $\und H$ explicitly given
as follows. Its linear space and algebra structure are unchanged, but are now
viewed as living in the quasitensor category of $H$-modules by the quantum
adjoint action.
This, and the modified coalgebra and antipode are\cite{Ma:bra}
\eqn{e3}{h\la b=\sum h\o b \ant h\t,\quad \und\Delta b=\sum b\o\ant\CR\ut\tens
\CR\uo\la b\t,\quad \und \ant b=\sum \CR\ut\ant(\CR\uo\la b)}
for all $h\in H$, $b\in\und H$. This gives the braided structures in terms of
those of our initial $H$. For our class $\CO$ we take the tautological $\und
H$-modules in the quasitensor category, where the action of $\und H$ on on
object $V$ is given by the same linear map as the action of $H$ that makes $V$
an object. In this case the braided-cocommutativity of $\und H$ then explicitly
takes the concrete form
\cite{Ma:bra},
\eqn{e4}{ \sum \Psi(b_{\und {(1)}}\tens Q\uo\la b_{\und {(2)}})Q\ut=\sum
b_{\und {(1)}}\tens b_{\und {(2)}}}
where $\und\Delta b=\sum b_{\und {(1)}}\tens b_{\und {(2)}}$ and
$Q=\CR_{21}\CR_{12}$. Here $Q\ut$ multiplies the result of $\Psi$ from the
right as $(1\tens Q\ut)$. This implies the condition in Figure~2(b) for all
$\CO$ and is essentially equivalent to it in the present context. $Q$
corresponds to the double-twist $\Psi_{V,B}\circ\Psi_{B,V}$ and so is absent in
the triangular case. For the example $H=U_q(g)$, we shall denote its associated
braided group by $\und H=BU_q(g)$.

Finally, we recall from \cite{Ma:eul}\cite{Ma:bg} that if $(A,\CR)$ is a dual
quantum group (with dual quasitriangular structure $\CR:A\tens A\to k$ obeying
some obvious axioms dual to (\ref{qtri})) then there is a corresponding braided
group $\und A$ (of function algebra type). As a coalgebra it coincides with
$A$, while the modified product and antipode take the form
\cite{Ma:eul}\cite{Ma:bg},
\eqn{e5}{a\und\cdot b=\sum a\t b\t \CR((\ant a\o)a\th,\ant b\o),\quad
\und{\ant}a=\sum \ant a\t\CR((\ant^2a\th)\ant a\o,a_{(4)})}
and there is a right adjoint coaction of $A$ so that this $\und A$ is a Hopf
algebra in the category of $A$-comodules.
If $A=H^*$ then the corresponding coadjoint action of $H$ on $\und A$ is $h\la
a=\sum a\t <h,(\ant a\o)a\th>$. There is also a right action $L^*$ of $H$ on
$A$ defined by $L^*_h(a)=\sum <h,a\o>a\t$. In terms of test-elements $g\in H$,
these are
\eqn{e6}{<h\la a,g>=<a,\sum (\ant h\o)gh\t>,\quad <L^*_h(a),g>=<a,hg>.}
The braided-commutativity of $\und A$ then takes the concrete form
\eqn{e7}{a\und\cdot b=\sum \und\cdot\circ\, Q\ut\la\Psi(L^*_{Q\uo}(a)\tens b)}
where $Q\ut$ acts on the first factor of the result of $\Psi$. This implies the
diagrammatic form of the braided-commutativity condition for all $\und
A$-comodules $\CO$ for which the coaction of $\und A$ is the tautological one.
At least in the finite-dimensional case, the two braided groups $\und A,\und H$
are dually paired in a certain sense (cf. \cite{LyuMa:bra}). Here we consider
both $\und A,\und H$ as living in the category of $H$-modules since every
(right) $A$-comodule defines a (left) $H$-module by dualization. For the
standard quantum function algebras $G_q$ dual to the $U_q(g)$, we denote the
associated braided groups of function algebra type by $BG_q$. The matrix
bialgebras $A(R)$\cite{FRT} mapping onto $G_q$ can also be converted in a
similar way and give the braided-matrices $B(R)$ introduced in
\cite{Ma:exa}\cite{Ma:eul}. They are braided-commutative bialgebras in the
category of $U_q(g)$-modules.

\section{BRAIDED GROUP OF $U_q(g)$ IN `FRT' FORM}

This section is devoted to a computation of the braided groups associated to
the quantum groups $U_q(g)$ in FRT form. The results will then be applied in
later sections.
Our first two technical results can in principle be motivated from category
theory\cite{Lyu:mod} along the lines of \cite{LyuMa:bra}. For our algebraic
purposes (and in the form we need now) we give a direct algebraic treatment.
The first establishes that $Q=\CR_{21}\CR_{12}$ defines a morphism $\und
A\to\und H$ in the category of $H$-modules.

\begin{propos}cf\cite{ResSem:mat} Let $H$ be a finite-dimensional quantum group
and $A=H^*$ its corresponding dual quantum group. Then the map
$Q:A\to H$ given by $Q(a)=\sum <Q\uo,a>Q\ut$ is an intertwiner for the adjoint
and coadjoint actions of $H$ above.
\end{propos}
\proof We compute the action of $H$ using the definitions above and standard
properties of quantum groups. $\CR'=\sum \CR'\uo\tens\CR'\ut$ denotes a second
identical copy of $\CR$. We have
\align{Q(h\la a)&&=\sum \CR'\uo\CR\ut<(\ant h\o)\CR'\ut\CR\uo h\t,a>\\
&&=\sum \CR'\uo \CR\ut h\th\ant h_{(4)}<(\ant h\o)\CR'\ut\CR\uo h\t,a>\\
&&=\sum \CR'\uo  h\t\CR\ut\ant h_{(4)}<(\ant h\o)\CR'\ut h\th \CR\uo,a>\\
&&=\sum h_{(3)}\CR'\uo  \CR\ut\ant h_{(4)}<(\ant h\o) h\t \CR'\ut\CR\uo,a>\\
&&=\sum h\o\CR'\uo  \CR\ut\ant h\t< \CR'\ut\CR\uo,a>=h\la Q(a).}
Here the third and fourth equalities use the intertwining property of the
quasitriangular structure $\CR$ for $\Delta$ with its opposite. \endproof

\begin{propos} The morphism $Q:\und A\to\und H$ established in Proposition~2.1
is a Hopf algebra homomorphism for the Hopf algebras $\und A,\und H$ in the
category of $H$-modules.  In particular, if $H$ is factorizable in the sense of
\cite{ResSem:mat}, we have $\und A\isom \und H$.
\end{propos}
\proof We compute with the product defined in (\ref{e5}) to give
\align{Q(a\und\cdot b)\nquad&&=\sum \CR\uo\CR'\ut<a\t
b\t,\CR\ut\CR'\uo><a\o,\CR''\uo><a\th,\CR'''\uo><b\o,(\ant\CR'''\ut)\CR''\ut>\\
&&= \sum \CR\uo\CR'\ut <a\tens b,\CR''\uo\CR\ut\o\CR'\uo\o\CR'''\uo\tens
(\ant\CR'''\ut)\CR''\ut\CR\ut\t\CR'\uo\t>}
Using the axioms for $\CR$ this is the evaluation with $a\tens b$ in the last
two factors of the element
$\sum X\uo\tens X\ut\CR'''\uo\tens (\ant \CR'''\ut)X\uth$ where
$X=\CR_{23}\CR_{13}\CR_{12}\CR_{21}\CR_{31}=\CR_{12}\CR_{13}\CR_{23}
\CR_{21}\CR_{31}$ by the quantum Yang-Baxter equations (QYBE) obeyed by $\CR$.
Writing $\CR_{13}=(\id\tens\ant)(\CR^{-1}_{13})$, combining the elements on
which $\ant$ acts and using the QYBE again for them we obtain (after
cancellations) the element $\CR_{12}\CR_{21}\CR_{13}\CR_{31}$. The pairing of
this with $a\tens b$ is just $Q(a)Q(b)$ as required. Next we compute with the
coproduct defined in (\ref{e3}) to give
\[\und\Delta Q(a)=\sum \CR\uo\o\CR'\ut\o\CR'''\ut\ant\CR''\ut\tens
\CR'''\uo\CR\uo\t\CR'\ut\t\CR''\uo<a,\CR\ut\CR'\uo>\]
This is evaluation of $a$ on the third factor of the element $\sum
X\uo\ant\CR'''\ut\tens \CR'''\uo X\ut\tens X\uth$ where
$X=\CR_{13}\CR_{23}\CR_{32}\CR_{31}\CR_{21}=\CR_{13}\CR_{23}\CR_{21}\CR_{31}
\CR_{32}$ by the QYBE. Writing
$\CR_{31}=(\ant\tens\id)(\CR^{-1}_{31})$, combining the arguments of $\ant$,
using the QYBE again and cancelling now gives
the element $\CR_{13}\CR_{31}\CR_{23}\CR_{32}$. The pairing of this with $a$
gives $(Q\tens Q)\circ\Delta a$ as required.
\endproof

Note that the notion of factorizability introduced in \cite{ResSem:mat} is
precisely that the linear map $Q:H^*\to H$ is a linear isomorphism. Our
propositions say that in the braided setting it becomes a Hopf algebra
isomorphism. Quantum doubles as well as the $U_q(g)$ at least for generic $q$
are known to be factorizable. For the latter we work as usual over formal
power-series in a parameter $\hbar$ where $q=e^{\hbar\over 2}$ as in
\cite{Dri}. In this case there is a suitable dual to play the role of $H^*$ in
the propositions above. The algebraic proofs above clearly extend to this
setting. We are therefore in a position to exploit the isomorphism $Q$ for
these Hopf algebras. To do this, it is convenient to work with the generators
in `FRT' form as follows.

Firstly, \cite{FRT:lie}(cf.\cite{Dri}) identified the duals of $U_q(g)$ as
quotients of bialgebras $A(R)$ for certain $R$-matrices $R\in M_n\tens M_n$
associated to the classical families of simple Lie algebras $g$. Here $A(R)$ is
the bialgebra with generators $t^i{}_j$ and relations $Rt_1t_2=t_2t_1R$ in
standard notations. \cite{FRT:lie} also showed how to recover $U_q(g)$ in some
form as an algebra $U(R)$ with matrix generators $l^\pm$ and various relations.
Among them are the matrix relations of the form
$l^\pm_1l^\pm_2R=Rl^\pm_2l^\pm_1$ and $l^-_1l^+_2R=Rl^+_2l^-_1$, as well as
many hidden relations among the $2n^2$ generators expressed in the form of an
ansatz for the $l^\pm$ in terms of the familiar generators for the $U_q(g)$. We
refer to this description of $U_q(g)$ by `matrix generators + ansatz'  as {\em
the `FRT' form} of $U_q(g)$. All our results below are
intended for the $U_q(g)$ in this form, and we rely on \cite{FRT:lie} for
details of their connection with other descriptions of $U_q(g)$ (this is known
at least for the classical families of Lie algebras $g$). In fact, if the
universal $\CR$ for $U_q(g)$ is known in any given set of generators, it can be
exploited to give the required ansatz easily according to
\eqn{e8}{ l^+=\sum \CR\uo<t,\CR\ut>,\quad l^-=\sum <t,\ant\CR\uo>\CR\ut,}
see \cite{Ma:dua} where this method was used to generate the ansatz for
$U_q(sl(3))$.

We also need the explicit description of the braided matrices $B(R)$ as
introduced in \cite{Ma:exa}. They are given by matrix generators $u^i{}_j$ and
certain matrix relations. The difference is that now, the $u^i{}_j$ span an
object in the quasitensor category of $U_q(g)$-modules. The action is $l^+_2\la
u_1=R^{-1}u_1R$ and $l^-_1\la u_2=Ru_2 R^{-1}$ \cite{Ma:exa}. The
braided-coproduct, braiding $\Psi$ and algebra relations take the
form\cite{Ma:exa}
\eqn{e9}{\und\Delta u^i{}_j=u^i{}_k\tens u^k{}_j,\quad \Psi(u^I\tens
u^K)=u^L\tens u^J\Psi^K{}_L{}^I{}_J,\quad
u^Iu^K=u^Lu^J\Psi'^K{}_L{}^I{}_J}
where the $u^I$ etc are $u^i{}_j$ written with multi-indices, matrix $\Psi$
comes from (\ref{e1}) and $\Psi'$ is a variant corresponding (in the quotient
Hopf algebra) to
the right hand side of (\ref{e7}). Explicitly, they are given by\cite{Ma:exa}
\eqn{psi}{\Psi^I{}_J{}^K{}_L=
R^{k_0}{}_a{}^d{}_{j_0} R^{-1}{}^a{}_{l_0}{}^{j_1}{}_b
R^{l_1}{}_c{}^b{}_{i_1} {\tilde R}^c{}_{k_1}{}^{i_0}{}_d}
\eqn{psiprime}{\Psi'{}^I{}_J{}^K{}_L=
R^{-1}{}^{d}{}_{j_0}{}^{k_0}{}_{a}
R^{j_1}{}_{b}{}^{a}{}_{l_0}R^{l_1}{}_c{}^b{}_{i_1} {\tilde
R}^c{}_{k_1}{}^{i_0}{}_d.}
where $\tilde R=((R^{t_2})^{-1})^{t_2}$ with ${}^{t_2}$ denoting transposition
in the second matrix factor. Another (more conventional) way to write the
relations of $B(R)$ is to move two of the $R$'s in $\Psi'$ to the left hand
side, in which case the equations become equivalently
\eqn{re}{R_{21}u_1R_{12}u_2=u_2 R_{21} u_1 R_{12}.}
These are quite similar to some equations in \cite{Skl:bou} as well as being
known in \cite{ResSem:cen}\cite{MorRes:com}\cite{AFS:hid} to be discussed
below. In our case they are nothing other than the braided-commutativity
(\ref{e7}) in the case of $B(R)$. The coproduct $\und \Delta$ extends to all of
$B(R)$ as a bialgebra in this quasitensor
category\cite{Ma:exa}. The construction is very general, but in the present
case (for the standard $R$ matrices) after further quotienting $B(R)$ by
`braided-determinant' type relations, one obtains $\und A=BG_q$, the braided
group corresponding to the
quantum group dual to $H=U_q(g)$ in the setting above.  We are now ready to
prove our main result of this section. It is a corollary of Proposition~2.2 in
the form over $\C[[\hbar]]$.

\begin{corol} Let $H=U_q(g)$ in FRT form\cite{FRT:lie} and $\und A=BG_q$ the
braided group of function algebra type as recalled above. Then the braided
group $\und H=BU_q(g)$ of enveloping algebra type corresponding to $U_q(g)$ has
the same algebra structure as $U_q(g)$ but the new coproduct implied by
\[ \und\Delta L^i{}_j=L^i{}_k\tens L^k{}_j,\quad{\rm where}\quad L=l^+\ant
l^-\]
The space spanned by the $L^i{}_j$ is a $U_q(g)$-module via $l^+_2\la
L_1=R^{-1}L_1R$ and $l^-_1\la L_2=RL_2 R^{-1}$ and the identification $L=u$
allows us to consider $BU_q(g)=BG_q$ as a self-dual braided group.
\end{corol}
\proof We compute the map $Q$ in Proposition~2.2 as
$Q(u^i{}_j)=\sum \CR\uo\CR'\ut<u^i{}_j,\CR\ut\CR'\uo>=\sum
\CR\uo\CR'\ut<u^i{}_k,\CR\ut><u^k{}_j,\CR'\uo>=\sum
\CR\uo\ant\CR'\ut<u^i{}_k,\CR\ut><u^k{}_j,\ant\CR'\uo>=l^+{}^i{}_k\ant
l^-{}^k{}_j=L^i{}_j$. The action of is $l^+{}^i{}_j\la
Q(u^k{}_l)=Q(l^+{}^i{}_j\la u^k{}_l)=R^{-1}{}^k{}_m{}^i{}_a L^m{}_n
R^n{}_l{}^a{}_j$ using Proposition~2.1 and the action on $u^k{}_l$ obtained
from (\ref{e6}) above and the pairing $<u,l^+>=R$ from \cite{FRT:lie}. Note
that the transmutation
procedure used to define these braided groups in \cite{Ma:exa} is such that we
can identify the generators $u$ with the generators $t$ of $G_q$ (but not their
products), and we have used this fact here. Similarly for the action of $l^-$.
(The action on
$B(R)$ above is similar, but in a more general setting). Because of
Proposition~2.1 this result on $L$ must coincide with the quantum adjoint
action as defined in (\ref{e3}). \endproof

These new generators $L$ for $U_q(g)$ are the ones in which the corresponding
$BU_q(g)$ becomes explicitly identified with
a quotient of $B(R)$ where the matrix coproduct is braided. Explicitly for
$U_q(sl_2)$  they are easily computed as
\eqn{e15}{L=\pmatrix{q^H&q^{-{1\over 2}}(q-q^{-1})q^{H\over 2}X_-\cr
q^{-{1\over 2}}(q-q^{-1})X_+q^{H\over 2}& q^{-H}+q^{-1}(q-q^{-1})^2X_+X_-}}
in the usual description for $U_q(sl_2)$ with the conventions
$[X_+,X_-]=(q^H-q^{-H})/(q-q^{-1})$ of \cite{Jim:dif}. These combinations have
been known in various contexts, notably \cite{MorRes:com}\cite{AFS:hid} and
cf\cite{ResSem:cen}\cite{Skl:bou}. There it is known that the relations
\eqn{e16}{R_{21}L_1R_{12}L_2=L_2 R_{21} L_1 R_{12}}
also describe the relations of $U_q(sl_2)$ (for example) given the ansatze for
$l^\pm$. We see from Corollary~2.3 that this is due to the factorizability of
$U_q(sl_2)$ and hence holds quite generally as an expression of the
braided-commutativity of $BG_q$ as in (\ref{e7}) carried over to $BU_q(g)$ via
the isomorphism  $BU_q(g)\isom BG_q$. Likewise, we learn that
$BG_q$ is braided-cocommutative since $BU_q(g)$ is. Explicitly, this
braided-cocommutativity takes the form
\eqn{e17}{\Psi(L^i{}_a\tens L^b{}_c)M^a{}_b{}^c{}_j=L^i_a\tens L^a{}_j,\quad
M^a{}_b{}^c{}_j=l^+{}^c{}_k(\ant^{-1} l^+{}^m{}_b)l^-{}^a{}_m\ant l^-{}^k{}_j.}
This is noted for completeness and is readily computed from (\ref{e4}) using
the same techniques as in the corollary above.

Also, in various other contexts it has been noted that such combinations as in
(\ref{e15}) are indeed fixed under the quantum adjoint action for $U_q(sl_2)$.
However, Corollary~2.3 ensures these desirable features hold quite generally,
giving us a fundamental ad-invariant subspace of $U_q(g)$ for all the standard
Lie algebras $g$. This suggests that this subspace should have properties
resembling some kind of `quantum Lie algebra'  (or `braided Lie algebra') for
$U_q(g)$. Recall that for an ordinary Lie algebra the vector space of $g$ is a
$g$-module by the adjoint action, and this action as a map
coincides with the Lie bracket. We can therefore likewise take the quantum
adjoint action on the space spanned by the $L^i{}_j$ as a `quantum Lie bracket'
or `braided Lie bracket'.

\begin{propos} Let $\CL\subset U_q(g)$ be the subspace spanned by the
$\{L^i{}_j\}$ in Corollary~2.3. We let $[\ ,\ ]:U_q(g)\tens U_q(g)\to U_q(g)$
be defined by $[h,g]=h\la g$ where $\la$ is the quantum adjoint action in
(\ref{e3}). This `quantum Lie bracket' enjoys the properties of closure and
`Jacobi' identities

(L0) $\quad [\xi,\eta]\in\CL$ for $\xi,\eta\in \CL$

(L1) $\quad [\xi,[\eta,\zeta]]=\sum [[\xi\o,\eta],[\xi\t,\zeta]]$

(L2) $\quad [[\xi,\eta],\zeta]=\sum [\xi\o,[\eta,[\ant\xi\t,\zeta]]]$

\noindent where $\Delta\xi=\sum \xi\o\tens\xi\t\in U_q(g)\tens U_q(g)$ is the
usual coproduct, and $\ant$ is the usual antipode. On the vectors $L^I$ where
$I=(i_0,i_1)$ we have
\[ [L^I,L^J]=c^{IJ}{}_K L^K,\quad c^{IJ}{}_K=R^{-1}{}^a{}_{i_1}{}^{j_0}{}_b
R^{-1}{}^b{}_{k_0}{}^{i_0}{}_c Q^c{}_{j_1}{}^{k_1}{}_a\]
where $Q=R_{21}R_{12}$.
\end{propos}
\proof The identity (L1) is an expression of the fact that the quantum adjoint
action is an intertwiner for itself, i.e.
in the general setting above it is a morphism $\und H\tens\und H\to\und H$ in
the category of $H$-modules. This a general feature of the braided groups $\und
H$ associated to $H$\cite{Ma:tra}. This, and the identity (L2) follow easily
from the definition of the quantum adjoint action in (\ref{e3}). For the
explicit form of the $c^{IJ}{}_K$ we use the same method as in Corollary~2.3 to
compute the action of $\ant l^-$ on $L$. It comes out as $(\ant l^-{}^i{}_j)\la
L^k{}_l=R^{-1}{}^a{}_j{}^k{}_m L^m{}_n R^i{}_a{}^n{}_l$. Using this and the
action of $l^+$ already given, we compute $[L^I,L^J]=L^I\la L^J$.
\endproof

Note that an easy computation gives $\Delta L^i{}_j=l^+{}^i{}_a\ant
l^-{}^b_j\tens L^a{}_b$ (as already noted in \cite{MorRes:com})  so that the
ordinary coproduct on the right hand side of (L1),(L2) does not have its image
in $\CL\tens\CL$. Hence these identities do not make sense for $\CL$ in
isolation from its quantum group $U_q(g)$. For this reason we do not formalize
these as axioms for an abstract Lie algebra. Nevertheless, if  we imagine that
$\xi$ is primitive as for a classical
lie algebra, i.e. $\Delta\xi=\xi\tens 1+1\tens\xi$ (and $\ant\xi=-\xi$) and
note that (unusually) $[1,\xi]=1\la\xi=\xi$, then the above do reduce to two
ordinary Jacobi identities for the two left hand sides. These two ordinary
Jacobi identities
imply on adding that $[\eta,[\xi,\zeta]]+[[\xi,\zeta],\eta]=0$, i.e. in the
semisimple case they imply antisymmetry. Thus (L1), (L2) together play the role
of one usual Jacobi identity and antisymmetry. On the other hand, not all
elements of $\CL$ are primitive, even as $q\to 1$, and indeed the properties of
the generators $\CL$ also express group-like as well as Lie-algebra like
features.

In addition, there are numerous other identities inherited from the structure
of $U_q(g)$ and its braided group, expressing the joint role of these as
`enveloping algebra' for $\CL$. These include
\align{[\xi\eta,\zeta]&\nquad=[\xi,[\eta,\zeta]],\quad & [\xi,\eta\zeta]=\sum
[\xi\o,\eta][\xi\t,\zeta]\\
\sum (\ant \xi\o)[\xi\t,\eta]&=\sum [\eta\o,\ant\xi]\eta\t,\quad &
[\xi,\eta]=\sum \xi\o\eta\ant\xi\t\\
(\sum [\xi\o,\ ]\tens [\xi\t,\ ])\und\Delta&\nquad=\und\Delta [\xi,\
]\qquad&(\id\tens\und\Delta)\und\Delta=(\und\Delta\tens\id)\und\Delta.}
If we imagine $\xi$ etc primitive as before, the second identity becomes
$[\xi,\eta\zeta]=[\xi,\eta]\zeta+\eta[\xi,\zeta]$, the third
becomes $[\xi,\eta]=[\eta,-\xi]$ and the fourth becomes
$[\xi,\eta]=\xi\eta-\eta\xi$. The last identity refers to the braided coproduct
of $BU_q(g)$, which restricts to $\und\Delta:\CL\to \CL\tens\CL$ as a
generalization of the coproduct in the universal enveloping algebra of a Lie
algebra. The identity corresponds to the fact there that if $\xi,\eta$ are
primitive, then $[\xi,\eta]$ is also primitive. The last identity is the
coassociativity inherited from that of $BU_q(g)$.
Thus, it is this $\und\Delta$ that preserves $\CL$ and extends to products of
the generators as a (braided) Hopf algebra.

Finally, in the remaining sections of the paper we will focus for concreteness
on the example of the above for $H=\usl$. There is a standard matrix $R$ for
this in the FRT approach. Then $B(R)=BM_q(2)$ (the braided matrices of $sl_2$
type) has the action on $u=\pmatrix{a&b\cr c&d}$ given by\cite{Ma:exa}
\eqn{e10}{ q^{H\over 2}\la\pmatrix{a&b\cr c&d}=\pmatrix{a&q^{-1}b\cr qc&d}}
\eqn{e11}{X_+\la\pmatrix{a&b\cr c&d}=\pmatrix{-q^{3/2}c&
-q^{1/2}(d-a)\cr0&q^{-{1/2}}c},\ X_-\la\pmatrix{a&b\cr c&d}=\pmatrix{q^{1/2}b&
0\cr q^{-{1/2}}(d-a)&-q^{-{3/2}}b}.}
The algebra relations are comparable to those of the quantum matrices $M_q(2)$
and come out as
\eqn{e12}{ba=q^2ab,\qquad ca=q^{-2}ac,\qquad da=ad,\qquad
bc=cb+(1-q^{-2})a(d-a)}
\eqn{e13}{db=bd+(1-q^{-2})ab,\qquad cd=dc+(1-q^{-2})ca}
The additional `braided-determinant' relation
\eqn{e14}{ad-q^2 cb=1} gives the braided group $BSL_q(2)$ (the braided group
version of $SL_q(2)$). This is a braided group of `function algebra' type and
is $\und A$ in the setting above when $H=U_q(sl_2)$. Our results above imply
that this can be
identified as $u=L$ with the braided group $\busl$ of enveloping algebra type.
Note that $BSL_q(2)$ has a bosonic central element $q^{-1}a+qd$ as explained in
\cite{Ma:exa}. It is the spin 0 generator in the identification $\CL=1\oplus 3$
where the remaining generators form a 3-dimensional spin 1 representation of
$U_q(sl_2)$. The element is bosonic in the sense that $\Psi((q^{-1}a+qd)\tens
f)=f\tens (q^{-1}a+qd)$ for all $f$ since the action of $\CR$ in (\ref{e1}) is
trivial. We see from the identification $u=L$ that this element is just the
quadratic Casimir of $\busl$.

In summary, we have shown that the generators $L=l^+\ant l^-$ of $U_q(g)$ are
convenient for the description of the corresponding braided group. This braided
group is at the same time the braided-cocommutative braided group of enveloping
algebra type consisting of $U_q(g)$ with a modified coproduct, and the
braided-commutative braided group of function algebra type dual to this.
Moreover, these generators $L$ exhibit a number of Lie-algebra type properties
inherited from these structures. In particular, $\busl=BSL_q(2)$ can be
identified.

\section{BRAIDED MATRIX STRUCTURE OF THE DEGENERATE SKLYANIN ALGEBRA}

The Sklyanin algebra was introduced in \cite{Skl:alg}\cite{Skl:rep} in
connection with an ansatz for the 8-vertex model. As an algebra it has been
extensively studied by ring-theorists for its remarkable properties, see
\cite{SmiSta:reg} and elsewhere. The algebra has four generators
$S_0,S_\alpha$, $\alpha=1,2,3$ and three structure constants $J_{12}, J_{23},
J_{31}$ subject to the constraint $J_{12}+J_{23}+J_{31}+J_{12}J_{23}J_{31}=0$
and the relations
\eqn{e18}{[S_0,S_\alpha]=\imath J_{\beta\gamma}\{S_\beta,S_\gamma\},\qquad
[S_\alpha,S_\beta]=\imath\{S_0,S_\gamma\}}
 where $\alpha,\beta,\gamma$ are from the set 1,2,3 in cyclic order and $\{\ ,\
\}$ denotes anticommutator.
There are two Casimir elements
\eqn{e19}{C_1=S_0^2+\sum_\alpha S_\alpha^2,\qquad C_2=\sum_\alpha S^2_\alpha
J_{\alpha}}
where $J_{\alpha\beta}=-{J_\alpha-J_\beta\over J_\gamma}$. The degenerate case
where (say) $J_{12}=0$ is well-known to be closely connected with the quantum
group $\usl$. We write $S_\pm=S_1\pm\imath S_2$ and $K_\pm=S_0\pm tS_3$ where
$t=\sqrt{J_{23}}$ (a fixed square root). Then the relations become
\eqn{e20}{[K_+,S_\pm]=\pm t\{K_+,S_\pm\},\quad [K_-,S_\pm]=\mp
t\{K_-,S_\pm\},\quad [K_+,K_-]=0,\quad [S_+,S_-]={1\over t}(K_+^2-K^2_-).}
Writing $q={1+t\over 1-t}$ and $Y_\pm=\h\sqrt{1-t^2}S_\pm$ we have
\eqn{e21}{[K_+,K_-]=0,\quad  K_+Y_\pm=q^{\pm 1}Y_\pm K_+,\quad
K_-Y_\pm=q^{\mp1}Y_\pm K_-,\quad [Y_+,Y_-]={K_+^2-K_-^2\over q-q^{-1}}}
while two independent linear combinations of the Casimir elements are (with
$J_1=J_2=1, J_3=1+J_{23}$)
\eqn{e22}{C_1-C_2=K_+K_-,\quad C\equiv
2({(1+t^2)\over(1-t^2)}C_1+C_2)=q^{-1}K_+^2+qK_-^2+(q-q^{-1})^2Y_+Y_-.}
Thus, with these changes of variables we see that the further quotient
$K_+K_-=1$ gives us the familiar algebra of $\usl$
(in Jimbo's conventions) and the combination $C$ shown becomes its familiar
quadratic Casimir element. $\usl$ is of course a Hopf algebra but, as far as is
known, the full degenerate Sklyanin algebra itself is not. Instead we have

\begin{theorem} The degenerate Sklyanin algebra as described is isomorphic to
the braided matrices $BM_q(2)$ of $sl_2$ type. Explicitly the generators
$u=\pmatrix{a&b\cr c&d}$ of the latter take the form
\[ \pmatrix{a&b\cr c&d}=\pmatrix{K^2_+& q^{-\h}(q-q^{-1})K_+Y_-\cr
q^{-\h}(q-q^{-1})Y_+K_+& K_-^2+q^{-1}(q-q^{-1})^2Y_+Y_-}
\]
and we allow $a, d-ca^{-1}b$ of $BM_q(2)$ to be invertible and have square
roots. Hence the degenerate Sklyanin algebra has the structure of a bialgebra
in the quasitensor category of $U_q(sl_2)$-modules.
The bosonic central elements $q^{-1}a+qd, ad-q^2cb$ of $BM_q(2)$ are explicitly
\[ q^{-1}a+qd=C,\quad ad-q^2cb=K_+^2K_-^2.\]

\end{theorem}
\proof Since the braided group $\busl$ has the same structure as an algebra as
the quantum group $\usl$, we know that the quotient of the degenerate Sklyanin
algebra by $K_+K_-=1$ is isomorphic to this also, and hence by Corollary 2.3
isomorphic to the braided group $BSL_q(2)$. But this is a quotient of $BM_q(2)$
by the braided-determinant $ad-q^2cb=1$, so we are motivated to make an ansatz
for $BM_q(2)$ in the form stated. The ansatz is then verified by explicit
computations which we leave to the reader. Clearly, the generators
$K_\pm^2,K_+Y_\pm$ can be recovered from the $a,b,c,d$. Thus, it is more
precisely this form of the Sklyanin algebra (rather than generators
$K_\pm,Y_\pm$) that is isomorphic to $BM_q(2)$. This is not, however, an
important distinction when we work over $\C[[\hbar]]$ with $K_\pm=q^{H_\pm\over
2}$ say with $q=e^{\hbar\over 2}$ (as for $\usl$). \endproof

We compute the braided structure in the Sklyanin algebra implied by this
theorem, as follows.

\begin{propos} The action of $\usl$ on the degenerate Sklyanin algebra as a
bialgebra in the category of $\usl$-modules is
explicitly
\[ q^{H\over 2}\la K_\pm=K_\pm,\quad q^{H\over 2}\la Y_\pm=q^{\pm 1}Y_\pm,\quad
X_\pm\la K_+=(1-q^{\pm 1})Y_\pm,\quad X_\pm\la K_-=(q^{\pm 1}-1)K_+^{-1}Y_\pm
K_-\]
\[ X_+\la Y_+=(1-q^{-1})Y_+^2 K_+^{-1},\quad X_+\la
Y_-=K_+^{-1}(Y_+Y_--qY_-Y_+)\]
\[ X_-\la Y_-=(1-q)Y_-^2 K_+^{-1},\quad X_-\la
Y_+=K_+^{-1}(Y_-Y_+-q^{-1}Y_+Y_-).\]
The degenerate Sklyanin algebra is invariant under this action in the sense
$h\la(ab)=\sum (h\o\la a)(h\t\la b)$ where $\Delta h=\sum h\o\tens h\t$ is the
usual coproduct of $\usl$.
\end{propos}
\proof This is determined from the form of the isomorphism in the preceding
theorem and the known action of $\usl$ on $BM_q(2)$ recalled in
(\ref{e10})-(\ref{e11}) from \cite{Ma:exa}. The action of $q^{H\over 2}$ is
easily determined first since this element is group-like so that $q^{H\over
2}\la(Y_+ K_+)=(q^{H\over 2}\la Y_+)(q^{H\over 2}\la K_+)$ etc. Similarly
$X_+\la (Y_+ K_+)=0$ according to (\ref{e11}), but
$X_+\la (Y_+ K_+)=(X_+\la Y_+)(q^{H\over 2}\la K_+)+ (q^{-{H\over 2}}\la
Y_+)(X_+\la K_+)=(X_+\la Y_+)K_++q^{-1}Y_+(X_+\la K_+)$ using the standard
coproduct of $X_+$. This determines $X_+\la Y_+$ once $X_+\la K_+$ is known.
This
is extracted from knowledge of $X_+\la K_+^2$ from (\ref{e11}) and a similar
computation. In the same way the action $X_+\la Y_-$ is extracted from from
$X_+\la b$ in (\ref{e11}). Finally, $X_+\la K_-$ can be extracted more easily
from $K_+^2K_-^2$ bosonic. The action of $X_-$ is then obtained by a symmetry
principle. These computations have been made and the resulting action verified
using the algebra package REDUCE. \endproof

In principle, we can similarly extract the form of the braiding $\Psi$ and the
braided coproduct $\und\Delta$ on the generators $K_\pm, Y_\pm$ from the
braiding and matrix coproduct $\und\Delta$ for $BM_q(2)$. From the theorem and
\cite{Ma:exa}, some examples of $\Psi,\und\Delta$ are
\eqn{e23}{\Psi(K_+^2\tens K_+Y_-)=K_+Y_-\tens K_+^2,\quad\Psi(Y_+K_+\tens
K_+Y_-)=q^{-2} K_+Y_-\tens Y_+K_+}
\eqn{e24}{\und\Delta K_+^2=K_+^2\tens K_+^2+q^{-1}(q-q^{-1})^2 K_+Y_-\tens Y_+
K_+}
\eqn{e25}{\und\Delta K_+Y_-=K_+^2\tens K_+Y_--q^{-2}K_+Y_-\tens K_-^2+
q^{-1}K_+Y_-\tens C}
\eqn{e26}{\und\Delta Y_+K_+=Y_+K_+\tens K_+^2-q^{-2}K_+^2\tens
Y_+K_++q^{-1}C\tens Y_+K_+}
However, in practice it is rather hard to proceed further to compute $\Psi$ and
$\und\Delta$ explicitly on $K_\pm,Y_\pm$ alone. This is because they do not
transform in a simple way among themselves under the action in Proposition~3.2
so that the braidings $\Psi(K_+\tens Y_+)$ etc as determined by $\CR$ in
(\ref{e1}) are given by infinite power-series rather than finite combinations
of the generators. This in turn means that the braided tensor product algebra
structure does not compute in closed form. Rather, we see that the generators
$K_+^2, q^{-\h}(q-q^{-1})Y_+K_+, q^{-\h}(q-q^{-1})K_+Y_-, C$ do transform among
themselves and are more convenient for the description of the braiding and
braided coproduct.

\section{BRAIDED STRUCTURE IN THE QUANTUM LORENTZ GROUP}

In this section we give a second application of Proposition~2.2 and
Corollary~2.3, this time to the algebraic structure of the quantum double of
the quantum groups $U_q(g)$. A physically interesting example of such a double,
namely the quantum double of $U_q(su(2)$ has been called the `quantum Lorentz
group' in \cite{PodWor:def} on the basis of a Hopf-$C^*$-algebraic  `quantum
Iwasawa decomposition'. Recall that the ordinary Lorentz group can be
identified (at the Lie algebra level) with $sl_2(\C)$
regarded as a real Lie algebra, i.e. the complexification of $su(2)$: the
quantum double $D(U_q(su(2))$ can likewise be regarded as a kind of
`complexification' of $U_q(su(2))$ as a Hopf $*$-algebra (or rather, in the
dual form as a  $C^*$-algebra). In fact, there are some quite general algebraic
arguments to arrive at this same conclusion, based on the author's
Yang-Baxter-theoretic proof of the Iwasawa decomposition for ordinary Lie
algebras appearing in \cite{Ma:mat}. We recall this first.

Let $u$ be a compact real form of a complex semisimple Lie algebra $g$. The
latter is the complexification of $u$ and forms a real Lie algebra of twice the
dimension: $g=\imath u\oplus u$ with the Lie bracket of $u$ extended linearly
to $g$. The Iwasawa decomposition states that there is a splitting $g=k\oplus
u$ as vector spaces into two sub-Lie algebras, with $k$ solvable. We observed
in \cite[Sec. 2]{Ma:mat} that this Lie algebra $k$ could be identified with the
Lie algebra structure on $u^\star$ associated to the Drinfeld-Jimbo solution
$r$
 of the Classical Yang-Baxter Equations (CYBE) as follows. Choosing a
Cartan-Weyl basis for $g$, the solution
$r\in g\tens g$ takes the form
\eqn{e27}{r=\sum_\lambda \sgn(\lambda){E_\lambda\tens E_{-\lambda}\over
K(E_\lambda,E_{-\lambda})}+K^{-1}}
where the sum is over root vectors $E_\lambda$ and $K$ denotes the Killing form
with inverse $K^{-1}$. In Drinfeld's
theory in \cite{Dri:ham} this defines a quasitriangular Lie bialgebra
$(g,\delta,r)$ where $\delta:g\to g\tens g$
is defined by $\delta\xi=\sum [\xi,r\uo]\tens r\ut+r\uo\tens [\xi,r\ut]$. Now,
just as every finite-dimensional Hopf algebra has a dual one built on the dual
linear space, every finite dimensional Lie bialgebra has a dual $g^*$. Its Lie
bracket is defined from $\delta$ by $<[\eta,\eta'],\xi>=<\eta\tens
\eta',\delta\xi>$ for $\eta,\eta'\in g^*$.
The key observation in \cite[Sec. 2]{Ma:mat} is that in our case, $r$ in
(\ref{e27}) has its first term (its
antisymmetric part) lying entirely in $\imath u\tens u$, while the second term
(its symmetric part) lies in $u\tens u$. Using this we showed that that the
subspace $u^\star\subset g^*$ defined by $u^\star=\imath K(u,\ )$ is fixed
under this Lie bracket on $g^*$. The coadjoint actions of $g$ on $g^*$ and
$g^*$ on $g$ restrict to mutual actions of
$u,u^\star$ on each other, and finally Drinfeld's Lie bialgebra double $D(g)$
built on the linear space of $g^{\rm *}\oplus g$\cite{Dri:ham} restricts to a
Lie bialgebra $u^{\star\rm op}\bowtie u$ built on $u^{\star} \oplus u\subset
g^*\oplus g$. For later use, the Lie algebra structure of $D(g)$ is explicitly
given by
\alignn{e28}{&&[\eta\oplus\xi,\eta'\oplus\xi']=([\eta',\eta]+ \sum \eta'\so
<\eta'\st,\xi>-\eta\so<\eta\st,\xi'>)\nonumber\\
&&\qquad\qquad\qquad\qquad\oplus ([\xi,\xi']+\sum
\xi\so<\eta',\xi\st>-\xi'\so<\eta,\xi'\st>)}
where $\delta\xi=\sum \xi\so\tens\xi\st$ etc is our explicit notation for the
cobrackets. The cobracket on $D(g)$ is the tensor product of those on $g$ and
$g^*$. We note that $u^{\star\rm op}$ etc denotes (in the present conventions)
$u^\star$ with its opposite (reversed) Lie bracket, while the notation
$u^{\star\rm op}\bowtie u$ derives from a general `double
semidirect sum' construction for a Lie algebra from a pair mutually acting on
each other in a compatible way (a `matched pair' of Lie algebras). We
introduced this notion in \cite[Sec. 4]{Ma:phy} where we showed that
$D(g)=g^{*\rm op}\bowtie g$ by the mutual coadjoint actions. Other authors have
also arrived at similar notions of Lie algebra
matched pairs, notably \cite{AmiKos:big}\cite{LuWei:poi}. Finally, there is an
isomorphism\cite[Sec. 2]{Ma:mat}
\eqn{e29}{  \phi: u^{\star \rm op}\bowtie u\isom g,\qquad \phi(\eta\oplus
\xi)=\sum r\uo<\eta,r\ut>+\xi.}
This isomorphism is our Yang-Baxter theoretic description of the Iwasawa
decomposition of $g$. Both the solvable Lie algebra $k=u^{\star\rm op}$ and the
decomposition itself are derived from the Drinfeld-Jimbo solution (\ref{e27}).

In
\cite[Sec. 3]{Ma:mat} we proceeded to construct a `matched pair' of Lie groups
$U,U^{\star\rm op}$ (say) by building from the mutual actions between
$u,u^{\star\rm op}$ a matching pair of gauge fields over $U^{\star\rm op}, U$
and using their parallel transport to exponentiate to global actions of the
groups. The resulting group double-semidirect product $U^{\star\rm op}\bowtie
U\isom G$ (where $G$ is the simply-connected Lie group of $g$) provided a new
constructive proof of the group Iwasawa decomposition.
In fact, the constructions were quite general, allowing for the exponentiation
of any Lie algebra splitting or `Manin triple' to a Lie group one provided some
technical criteria were satisfied (this was the main result of \cite{Ma:mat}).

We can however, go in another direction, namely to deform to the the quantum
group setting. Here the relevant notion, the `double cross product' of mutually
acting Hopf algebras (matched pairs of Hopf algebras) was introduced in
\cite[Sec. 3]{Ma:phy}. Every factorization of a Hopf algebra into sub-Hopf
algebras as defined in \cite{Ma:phy} can be reconstructed from its factors by
this double cross product construction. Once again, we showed that Drinfeld's
quantum
double $D(H)$, where $H$ is any (say, finite-dimensional) Hopf algebra, is
simply a Hopf algebra double cross product $D(H)= H^{*\rm op}\bowtie H$, this
time by mutual quantum coadjoint actions.  In the conventions that we need
below,
 $H^{*\rm op}$ denotes $H^*$ with the opposite product (more usually, one takes
here the opposite coproduct\cite{Dri}, but this is isomorphic via the antipode
$\ant$.  We have not discussed in \cite{Ma:phy} the question of real forms
($*$-structures) but it is clear that just as $u^{\star\rm op}\bowtie u\isom
\imath u\oplus u$ in (\ref{e29}) is a real form
of $g^{*\rm op}\bowtie g=D(g)$, so $D(U_q(g))$ should be regarded as a Hopf
algebra whose real form is the complexification of the real form $U_q(u)$ of
$U_q(g)$. Our algebraic results below are thus a further step towards an
Iwasawa decomposition theorem for quantum groups. We will obtain an analogue of
the formula (\ref{e29}) with the role of $r$ played by the universal R-matrix
of the quantum group.

Finally, working over $\C$ as we do brings out some further structure not
visible over $\R$. Namely, when regarding $g=\imath u\oplus u$, there is a
sense in which the elements of $\imath u$ (the pure boosts in the case of the
Lorentz group) are acted upon by the elements of $u$ (the rotations). I.e., the
Lorentz group Lie algebra (in addition to its numerous other descriptions) has
the flavour of a semidirect sum where the rotations act on the boosts by
commutation. On the other
hand, this cannot be literally so since the boosts do not close under
commutation. In fact, we will  see that $g=\imath u\oplus u$ can be embedded in
a natural way in a semidirect sum $g\cocross g$, with the `boosts' acted upon
by `rotations'. This is related to our result $D(g)\isom g\cocross g$ below. We
will also see the latter result in the quantum case for $D(H)$.

We begin (as in Section~2) with a general result for finite-dimensional quantum
groups $H$.  Its origins are in a result in \cite{Ma:dou} that the quantum
double $D(H)$ in this case (when $H$ is quasitriangular) has the structure of a
semidirect product. The result was obtained before the notion of braided groups
had been introduced. We need the following
more explicit variant. In the conventions that we need, we build the quantum
double $D(H)$ on $H^*\tens H$
as follows.  The coproduct, counit and unit for $D(H)$ are the tensor product
ones while the product of $D(H)$ comes out as
\eqn{e30}{ (a\tens h)(b\tens g)=\sum b\t a\tens h\t g<\ant h\o,b\o><h\th,b\th>}
for $h,g$ in $H$ and $a,b$ in $H^*$. Its antipode is $\ant(a\tens h)=(1\tens
\ant h)(\ant^{-1}a\tens 1)$. We have

\begin{propos} Let $H$ be a finite-dimensional quasitriangular Hopf algebra
with quantum double $D(H)$, $A=H^*$ and $\und A$ the associated braided group
of function algebra type. Then $D(H)\isom \und A\cocross H$ as a semidirect
product by the coadjoint action of $H$ on $\und A$ and as a semidirect
coproduct with the $H$ coaction induced by $\CR:A\to H$. Explicitly, the
semidirect product and coproduct on $\und A\cocross H$ are
\[ (a\tens h)(b\tens g)=\sum a\und\cdot(h\o\la b)\tens h\t g,\quad \Delta
(a\tens h)=\sum a\o\tens \CR\ut h\o\tens \CR\uo\la a\t\tens h\t\]
and the required isomorphism $\theta:\und A\cocross H\to D(H)$ is
$\theta(a\tens h)=\sum a\o<\CR\uo,a\t>\tens \CR\ut h$.
\end{propos}
\proof An abstract category-theoretic explanation of this result has recently
been given in \cite{Ma:cat}. However, for our present purposes we need a
completely explicit algebraic version as stated. Firstly, let us note that if
$\beta:\und A\to H\tens \und A$ is any left comodule structure respecting $\und
A$ as a coalgebra, the semidirect coproduct coalgebra is $\Delta (a\tens h)=
\sum a\o\tens a\t\bo h\o\tens a\t\bt\tens h\t$ where $\beta(a)=\sum a\bo\tens
a\bt$ denotes $\beta$ explicitly. This is a standard construction dual to the
equally standard semidirect product algebra construction stated. In the present
case the action is the one on $\und A$ in (\ref{e6}) and the coaction is
$\beta(a)=\sum \CR\ut\tens \CR\uo\la a$ (this is the way that any action of a
quasitriangular Hopf algebra is converted by $\CR$ to a coaction\cite{Ma:dou}).
We now
verify that $\theta$ is an isomorphism of coalgebras by computing
\align{&&\nqquad(\theta\tens\theta)\Delta (a\tens h)\\
&&=\sum a\o<\CR\uo,a\t>\tens \CR\ut\CR''\ut h\o\tens
a_{(4)}<\CR'\uo,a_{(5)}>\tens\CR'\ut h\t<\CR''\uo,(\ant a\th)a_{(6)}>\\
&&=\sum a\o<\CR\uo\o,a\t>\tens \CR\ut h\o\tens
a_{(4)}<\CR'\uo,a_{(5)}>\tens\CR'\ut h\t<\CR\uo\t,(\ant a\th)a_{(6)}>\\
&&=\sum a\o<\CR\uo,a_{(4)}>\tens \CR\ut h\o\tens a\t<\CR'\uo,a\th>\tens\CR'\ut
h\t\\
&&=\sum a\o\tens \CR\ut h\o\tens a\t\tens\CR'\ut h\t<\CR'\uo\CR\uo,a\th>\\
&&=\sum a\o\tens \CR\ut\o h\o\tens a\t\tens\CR\ut\t
h\t<\CR\uo,a\th>=\Delta_{A\tens H}\theta(a\tens h).}
For the first equality we used the definitions of $\theta$ and the stated
coproduct on $\und A\cocross H$. For the second and fifth we used the axioms of
the quasitriangular structure $\CR$, for the third we used the duality between
$H$ and $A$ and the antipode axioms. We verify that $\theta$ is an isomorphism
of algebras by computing
\align{&&\nqquad\theta((a\tens h)(b\tens g))\\
&&=\theta(\sum (h\o\la b)\bo a\o\tens h\t g<\CR,a\t\tens (h\o\la b)\bt>)\\
&&=\sum \theta((\CR\ut h\o\la b) a\o \tens h\t g)<\CR\uo, a\t>\\
&&=\sum (\CR\ut h\o\la b)\o a\o\tens \CR'\ut h\t g <\CR'\uo,(\CR\ut h\o\la b)\t
a\t><\CR\uo, a\th>\\
&&=\sum (\CR\ut h\o\la b)\o a\o\tens \CR'\ut h\t g
<\CR'\uo\t\CR\uo,a\t><\CR'\uo\o, (\CR\ut h\o\la b)\t>\\
&&=\sum  b\t a\o\tens \CR'\ut h\t g <\CR'\uo\t\CR\uo,a\t><\CR'\uo\o, b\th>
<\CR\ut h\o, (\ant b\o)b_{(4)}>\\
&&=\sum  b\t a\o\tens \CR'\ut\CR''\ut h\t g <\CR''\uo\CR\uo,a\t><\CR'\uo, b\th>
<\CR\ut h\o, (\ant b\o)b_{(4)}>\\
&&=\sum  b\t a\o\tens \CR'\ut\CR''\ut h\th g \\
&&\qquad\qquad\qquad\qquad<\CR''\uo\CR'''\uo\CR\uo,a\t> <\CR\ut h\o, \ant
b\o><\CR'\uo,b\th><\CR'''\ut h\t,b_{(4)}>\\
&&=\sum  b\t a\o\tens \CR'\ut\CR''\ut h\th g \\
&&\qquad\qquad\qquad\qquad<\CR''\uo\CR'''\uo\CR\uo,a\t> <\CR\ut h\o, \ant
b\o><\CR'\uo\CR'''\ut h\t,b\th>.}
The first equality uses (\ref{e5}) in the definition of the semidirect product
algebra structure on $\und A\cocross H$, writing (\ref{e5}) explicitly in terms
of the right adjoint coaction corresponding to the coadjoint action $\la$ in
(\ref{e6}). That $\la$ is an action then  gives the second equality. The fifth
equality uses the coadjoint coaction
again, in explicit form $b\mapsto\sum b\t\tens (\ant b\o)b\th$. The sixth and
seventh equalities use the axioms for $\CR$.  On the other side we compute with
the product $\cdot$ in $D(H)$ from (\ref{e27}), the expression
\align{&&\nqquad\theta(a\tens h)\cdot\theta(b\tens g)=\sum (a\o\tens\CR\ut
h)\cdot (b\o\tens \CR'\ut g)<\CR\uo,a\t><\CR'\uo,b\t>\\
&&=\sum b\t a\o\tens\CR\ut\t h\t \CR'\ut g\\
&&\qquad\qquad\qquad\qquad<\CR\uo,a\t><\CR'\uo,b_{(4)}><\CR\ut\o h\o,\ant
b\o><\CR\ut\th h\th,b\th>\\
&&=\sum b\t a\o\tens\CR''\ut h\t \CR'\ut g\\
&&\qquad\qquad\qquad\qquad<\CR'''\uo\CR''\uo\CR\uo,a\t><\CR\ut h\o,\ant
b\o><\CR'''\ut h\th\CR'\uo,b\th>\\
&&=\sum b\t a\o\tens\CR''\ut \CR'\ut h\th g\\
&&\qquad\qquad\qquad\qquad<\CR'''\uo\CR''\uo\CR\uo,a\t><\CR\ut h\o,\ant
b\o><\CR'''\ut\CR'\uo h\t,b\th>.}
For the last two equalities we used the axioms for $\CR$. Comparing these two
results we see that $\theta(a\tens h)\cdot\theta(b\tens g)=\theta((a\tens
h)(b\tens g))$ in virtue of the QYBE in the form
$\CR'''_{23}\CR''_{21}\CR'_{31}=\CR'_{31}\CR''_{21}\CR'''_{23}$. The other
facts such as the unit and counit are easy. \endproof

\begin{corol} If $H$ is a finite-dimensional factorizable quantum group then
there is an isomorphism $\phi=Q\circ\theta^{-1}:D(H)\isom \und H\cocross H$
where the semidirect product is by the
quantum adjoint action $\la$ of $H$ on $\und H$. Explicitly, $\phi(a\tens h)
=\sum Q(a\o)<\CR^{-(1)},a\t>\tens \CR^{-(2)}h$.\end{corol}

This is immediate from Proposition~4.1 and Proposition~2.2 in Section~2. Once
again, we can apply this more generally if we have a suitable notion of dual
Hopf algebra. In the setting where $H=U_q(g)$ we have

\begin{corol} Let $U_q(g)$ be in `FRT' form as in Corollary~2.3. Let $t$ be the
matrix generator of $G_q$, $L=l^+\ant l^-$ that of $U_q(g)$ and $M=m^+\ant m_-$
that of $BU_q(g)$ (it is the same algebra). Under the isomorphism
$\phi:D(H)\isom BU_q(g)\cocross U_q(g)$ the element $t^i{}_j\tens L^k{}_l$
corresponds to $M^i{}_a\tens l^-{}^a{}_jL^k{}_l$.

Explicitly, the structure of $BU_q(g)\cocross U_q(g)$ consists of the two
copies of $U_q(g)$ generated by $L,M$ as subalgebras with cross relations and
coproduct
\[ R_{12}l_2^+M_1=M_1 R_{12}l_2^+,\ R_{21}^{-1}l_2^- M_1=M_1R_{21}^{-1}l_2^-,\
\Delta l^\pm=l^\pm\tens l^\pm,\ \Delta M=(\sum M\CR\ut\tens \CR\uo
M)\CR^{-1}_{21}\]
where $\CR\in U_q(g)\tens U_q(g)$ as generated by $L$.\end{corol}
\proof This follows at once from the explicit form of $\phi$ in Corollary~4.2
and (\ref{e8}). The explicit form of the cross relations is nothing other than
the quantum adjoint action $l^\pm\la M$ computed as explained in the proof of
Corollary~2.3. The coproduct structure is the one in Proposition~4.1 computed
in the present case with the aid of the semidirect product algebra structure
and (\ref{qtri}). \endproof

These results show in particular that the quantum Lorentz group can be put into
semidirect product form, $D(\usl)=\busl\cocross\usl$. The price we pay for
keeping this more familiar semidirect product form is that the algebra
containing the `boosts' must be treated with braid statistics as the braided
group $\busl$ (as an algebra, it coincides with $\usl$). It is not any kind of
ordinary Hopf algebra, but a braided one in the category of $\usl$-modules. The
quantum
`rotations' $\usl$ can remain unchanged as an ordinary Hopf algebra and the
result (as a quantum double) is again a factorizable ordinary Hopf algebra. In
order to better understand this interpretation of Corollary~4.3, we pause now
to compute its classical meaning at the level of Lie bialgebras.

\begin{theorem} Let $(g,\delta,r)$ be a quasitriangular Lie bialgebra with
non-degenerate ad-invariant symmetric part $r_+$ of $r$,  and $D(g)$ its
double. Then there is an isomorphism $\phi: D(g)\isom g\cocross g$, where
$g\cocross g$ is the semidirect sum by the adjoint action of $g$ on itself.
Explicitly, it is given by
\[ \phi(\eta\oplus\xi)=2 r_+(\eta)\oplus(\xi-\sum <\eta,r\uo>r\ut).\]
Here we view $r_+$ as a linear map $r_+:g^*\to g$.
\end{theorem}
\proof The proof is by direct computation from (\ref{e28}) using the maps
shown. An introduction to the necessary methods of Lie bialgebras is to be
found in \cite[Sec. 1]{Ma:phy}. Firstly, let us recall that the structure of a
semidirect sum by (in our case) the adjoint action means
\eqn{e31}{[\zeta\oplus\xi,\zeta'\oplus\xi']=([\zeta,\zeta']+\alpha_\xi(\zeta')
-\alpha_{\xi'}(\zeta))\oplus [\xi,\xi'];\quad\alpha_\xi(\zeta)=[\xi,\zeta]}
for $\zeta\oplus\xi,\zeta'\oplus \xi'\in g\cocross g$. Using this, we have
\[[\phi(\eta\oplus\xi),\phi(\eta'\oplus\xi')]=
([2r_+(\eta),2r_+(\eta')]+[\xi-r(\eta),2r_+(\eta')]-[\xi'-r(\eta'),2r_+(\eta)])
\oplus [\xi-r(\eta),\xi'-r(\eta')]\]
Where we have written $\sum <\eta,r\uo>r\ut=r(\eta)$. On the other side we
compute using (\ref{e28}),
\align{&&\nqquad\phi([\eta\oplus\xi,\eta'\oplus\xi'])
=2r_+([\eta',\eta]+\eta'\so<\eta'\st,\xi>-\eta\so<\eta\st,\xi'>
)\\
&&\oplus([\xi,\xi']+\xi\so<\eta',\xi\st>-\xi'\so<\eta,\xi'\st>-r([\eta',\eta])
-r(\eta'\so)<\xi,\eta'\st>+r(\eta\so)<\xi',\eta\st>.}
For brevity, we omit summation signs. These two displayed expressions are equal
as follows. Firstly, $r([\eta',\eta])=
[r(\eta'),r(\eta)]$ is simply the CYBE when $r$ is viewed as a map $g^*\to g$
as we do here and the bracket $[\eta',\eta]$ is the one on $g^*$ also defined
by $r$ via $\delta$ on $g$ (this is equivalent to the more usual form
$[r_{12},r_{13}]+[r_{12},r_{23}+[r_{13},r_{23}]=0$ of the CYBE).  Secondly, We
have $\xi\so<\eta',\xi\st>-r(\eta'\so)<\xi,\eta'\st>
=-\xi\so<\eta',\xi\st>-<\eta',[\xi,r\uo]>r\ut=-[\xi,r\ut]<r\uo,\eta'>
=-[\xi,r(\eta')]$. Here we used antisymmetry of $\delta\xi$ and then its
explicit
form $[\xi,r\uo]\tens r\ut+r\uo\tens [\xi,r\ut]$. Thirdly, we have
$r_+\uo<\eta',[r_+\ut,\xi]>=-[r_+\ut,\xi]<\eta',r_+\uo>=[\xi,r_+(\eta')]$ by
ad-invariance of $r_+$. Fourthly we compute
\align{r_+([\eta',\eta])&&=<\delta r_+\uo,\eta'\tens\eta>r_+\ut\\
&&=<[r_+\uo,r\uo]\tens r\ut+r\uo\tens [r_+\uo,r\ut],\eta'\tens\eta>r_+\ut\\
&&=<-[r_+\uo,r\ut]\tens r\uo+2[r_+\uo,r_+'\uo]\tens
r_+'\ut,\eta'\tens\eta>r_+\ut+[r(\eta'),r_+(\eta)]\\
&&=[r(\eta'),r_+(\eta)]-[r(\eta),r_+(\eta')]+2[r_+(\eta),r_+(\eta')]}
where we used the definitions of the bracket in $g^*$ in terms of $\delta$ on
$g$. For last term in the third equality we used the previous (third)
observation applied to $\xi=r(\eta')$. Similarly for the final result. After
these four
observations we see that the expressions for
$[\phi(\eta\oplus\xi),\phi(\eta'\oplus\xi')]$ and
$\phi([\eta\oplus\xi,\eta'\oplus\xi'])$ coincide, i.e. $\phi$ is a Lie algebra
homomorphism. \endproof

Recall that the notion of a Lie bialgebra was introduced by Drinfeld as the
infinitesimal notion of a Hopf algebra. Thus, if we write $\Delta \xi=\xi\tens
1+1\tens\xi +\h\delta\xi +\cdots$ where we consider $\delta$ a deformation of
order $\hbar$, then to lowest order in $\hbar$ the formulae (\ref{e30}) reduce
to the structure of the Lie bialgebra double $D(g)$ in (\ref{e28}). The
formulae for the preceding theorem were obtained in the same way from
Corollary~4.2 with $\CR=1+r+\cdots$ where $r$ is also considered of order
$\hbar$. For example
$Q=\CR_{21}\CR_{12}=1+r_{21}+r_{12}+\cdots$. Thus, the notion of
`factorizability' of quantum groups is, at the level of Lie bialgebras just the
notion that the ad-invariant symmetric part if $r$ be non-degenerate. Thus the
role of $Q:\und A\to \und H$ in Proposition~2.1 is precisely played by
$2r_+:g^*\to g$.  For the solution (\ref{e27}) this is given by
twice the inverse Killing form $K^{-1}:g^*\to g$. Note that the isomorphism
$\phi$ in Theorem~4.4 works also at the level of cobrackets in the form
$D(g)\isom\und g\cocross g$ where $\und g$ denotes the Lie algebra $g$ equipped
with a certain modified (`braided') cobracket $\und\delta$. Finally, the
isomorphism $\phi$ clearly resembles the Iwasawa decomposition (\ref{e29}),
with $\phi$ in Corollaries~4.2 and 4.3
as quantum analogues. Indeed,

\begin{corol} Let $g=\imath u\oplus u$ denote the complexification of a real
Lie algebra $u$. There is a canonical embedding $g\subset g\cocross g$ such
that the restriction of $\phi$ in Theorem~4.4 to $u^{\star\rm op}\bowtie
u\subset D(g)$ recovers the Iwasawa decomposition (\ref{e29}). It is
\[ g\subset g\cocross g,\quad \xi_1+\imath\xi_2\mapsto 2\imath\xi_2\oplus
(\xi_1-\imath \xi_2).\]
\end{corol}
\proof This is obtained by computing $\phi^{-1}(\xi_1+\imath\xi_2)=\imath
K(\xi_2,\ )\oplus(\xi_1+\sum \imath K(r_-\uo
,\xi_2)r_-\ut)$ as the inverse of the Iwasawa decomposition (\ref{e29}). We
then apply to this the map $\phi$ in Theorem~4.4 which, for the Drinfeld-Jimbo
solution (\ref{e27}) takes the form
\[ \phi(\eta\oplus\xi)=2 K^{-1}(\eta)\oplus (\xi-K^{-1}(\eta)-\sum
<\eta,r_-\uo>r_-\ut).\]
Applying this gives  $2\imath\xi_2\oplus (\xi_1-\imath \xi_2)$ as stated. Note
that once found, one can easily verify this embedding $g\subset g\cocross g$
by elementary means (it holds for any real Lie algebra $u$), and hence regard
the Iwasawa decomposition as merely the restriction of $\phi$ in Theorem~4.4 to
a `real part'. This is the reason we have denoted both maps by $\phi$.
We leave the direct proof that the stated embedding $g\subset g\cocross g$ is a
Lie algebra homomorphism to the reader. $g$ has the Lie algebra structure of
$u$ extended linearly, while $g\cocross g$ has the semidirect product one in
(\ref{e31}). \endproof

We are now in a position to make precise our remarks about the Lorentz group.
We take $u=su(2)$ and $g=o(1,3)=sl_2(\C)=\imath u\oplus u$. Physically, the
real $u$ has compact generators $J_i$ (say) of a\eqn{e32}{
[J_i,J_j]=\eps_{ijk}J_k,\quad [J_i,K_j]=\eps_{ijk}K_k,\quad
[K_i,K_j]=-\eps_{ijk}J_k.}
The semidirect sum $o(1,3)\cocross o(1,3)$ is built on $o(1,3)\oplus o(1,3)$ in
the usual way
by (\ref{e31}) and the embedding in Corollary~4.5 is by
\eqn{e33}{J_i\mapsto 0\oplus J_i, \qquad K_i\mapsto 2K_i\oplus(-K_i).}
Thus it embeds rotations as rotations in the second $o(1,3)$ and boosts as
boosts in the first $o(1,3)$ along with a `compensating' negative boost in the
second. Apart from this compensation, the embedded boosts are acted upon by the
rotations as part of the semidirect sum. This unusual embedding corresponds to
the `real' part $su(2)^{\star\rm op}
\bowtie su(2)\subset D(o(1,3))$ where $o(1,3)\isom su(2)^{\star\rm op}\bowtie
su(2)$ is the Iwasawa decomposition and $D(o(1,3))\isom o(1,3)\cocross o(1,3)$
is from Theorem~4.4. This embedding is of course quite distinct from the more
usual identification $o(1,3)_\C\isom sl_2(\C)\oplus sl_2(\C)$ made in physics.
The latter is special while our embedding, although less familiar, is canonical
in the sense that a corresponding one holds for all complexifications $g$.

This completes our study of the algebraic structure of Drinfeld's quantum
double for the case of $D(U_q(g))$. We gave the general theory, the quantum
group setting and the classical limit. We return now to the general setting and
note that we can identify the semidirect structure in Proposition~4.1 and
Corollary~4.2 as examples of algebraic `bosonization'\cite[Sec. 4]{Ma:bos}.
There we show that if $B$ is any Hopf algebra in the quasitensor category of
modules of a quantum
group $H$ then the semidirect product and coproduct $B\cocross H$ along the
lines of Proposition~4.1 is an ordinary Hopf algebra. Thus our result is that
for a quantum group $H$, the Drinfeld quantum double $D(H)$ is the bosonization
of $\und A$ or (in the factorizable case) of $\und H$.

We can also use some more of the general theory in \cite{Ma:tra} to go further
and transmute the Hopf algebra $D(H)$ itself into a braided one. The general
transmutation principle in \cite{Ma:tra} asserts that if $H\to H_1$ is a Hopf
algebra map between ordinary Hopf algebras (with $H$ a quantum group, i.e. with
universal R-matrix) then $H_1$ with the same algebra acquires the additional
structure of a Hopf algebra in the quasitensor category of $H$-modules, denoted
$B(H,H_1)$.

\begin{propos} Let $H$ be a finite-dimensional quantum group and $D(H)$ its
double. The transmutation $B(H,D(H))$ of $D(H)$ into a Hopf algebra in the
quasitensor category of $H$-modules is $B(H,D(H))$ $\isom \und A\cocross \und
H$, a semidirect product algebra (with tensor product coalgebra) in the
category. If $H$ is factorizable then $B(H,D(H))\isom \und H\cocross \und H$.
\end{propos}
\proof This follows from the identification in Proposition~4.1 of $D(H)$ as the
result of bosonization of $\und A$. In general, the bosonization theorem of
\cite{Ma:bos} proceeds by forming the tautological semidirect product
$B\cocross \und H$ of the Hopf algebra $B$ in the category by the braided group
$\und H$ acting by the same action of $H$ by which $B$ is an object in the
category. The construction works because $\und H$ really behaves as a `group'
in the sense that it is braided-cocommutative. The resulting cross product is
then identified in \cite{Ma:bos} as the result of the transmutation of an
ordinary Hopf algebra $H_1$ by a map $H\to H_1$, $H_1$ being the bosonization.
We work the argument in reverse. \endproof

The semidirect product in the preceding proposition is not any complicated
quantum group semidirect product as in Proposition~4.1 and Corollary~4.2: it is
precisely the semidirect product by a group in the usual way (with no twisting
of the coproduct) except that all objects are treated with braid statistics. It
is precisely like the semidirect product by a super-group for example, with the
role of $\pm1$ played by $\Psi$. We have,

\begin{corol} Let $H=U_q(g)$ in `FRT' form. Then $\und H\cocross \und H$
explicitly has the structure on $\und H\tens\und H$ as follows. We denote
$L=1\tens L$ and $M=M\tens 1$ for the generators of the two copies of $\und H$.
These are embedded as sub-Hopf algebras in the quasitensor category of
$H$-modules, with the cross relations and coproduct
\[ L^k{}_l M^i{}_j=L^k{}_m\la\Psi(L^m{}_l\tens M^i{}_j),\quad
\und\Delta(M^i{}_j\tens L^k{}_l)=M^i{}_m\tens \Psi(M^m{}_j\tens L^k{}_n)\tens
L^n{}_l.\]
\end{corol}
\proof This is immediate the from the definition of cross products in
quasitensor categories studied in \cite[Sec. 2]{Ma:bos}. The element $L^k{}_m$
acts on the left factor of the result of $\Psi$ by the braided group adjoint
action $\la$ which, in the present case, coincides as a linear  map with the
quantum adjoint action (\ref{e3}). \endproof

Thus, if we are prepared to work with the `quantum Lorentz group' entirely in
the quasitensor category of $\usl$-modules, then it
takes a very natural form as simply the semidirect product of two identical
copies of the braided group $\busl$ with one of them (containing the `braided
boosts') acted upon (via the adjoint action) by the other (the `braided
rotations').

This algebraic analysis of the structure of the quantum Lorentz group (for
example) raises an interesting problem: what is the right notion of
$*$-structure for Hopf algebras in quasitensor categories? This is not a simple
problem since for a worthwhile notion of $*$-structure one has to consider also
what should be a `braided Hilbert space' and the corresponding adjoint
operation, before the right notion of unitarity etc in this braided setting is
found. (The situation is
complicated by the fact that $\Psi^2\ne\id$). We can hope that there can be
found such a notion such that we can compute quantum and braided real forms of
the above results along the lines of (\ref{e27}), and perhaps making contact
with the approach of \cite{PodWor:def}. See also \cite{Pod:com}. This is a
direction for further work.

\appendix
\section{BRAIDED GROUPS OF QUANTUM DOUBLES}

In this section we study one of the simplest examples of a factorizable Hopf
algebra, namely the braided group of the quantum double $D(H)$ of a general
Hopf algebra $H$. It is useful to see how some of the general theory of braided
groups, such as the self-duality in Proposition~2.2 looks in this case. This is
even more transparent for the simplest
ase of all, namely $D(G)$, where $G$ is a finite group. Here the self-duality
appears like the self-duality of $\R$ as expressed by $C_0(\R)\isom C^*(\R)$
(the Fourier convolution theorem). Moreover, quantum doubles $D(G)$ (and
quasi-Hopf algebra extensions of them)
have been identified in certain non-Abelian anyonic systems and in the context
of orbifold-based rational conformal field theories\cite{DPR:qua}. In both
cases one can work equally well with the corresponding braided group. The
category of $D(G)$-modules in which the braided-groups live is also an
interesting one and includes the category of crossed $G$-sets as introduced by
Whitehead\cite{Whi:com}. By developing our results for this simple discrete
quantum group $D(G)$ we hope to provide an antidote to the more abstract
results in the text.

We begin however, with the braided version of general $D(H)$, before passing to
our example. Thus $H$ denotes an arbitrary finite-dimensional Hopf algebra. The
structure of $D(H)=H^{*\rm op}\bowtie H$ was recalled in (\ref{e30}) above. It
contains both $H$ and $H^{*\rm op}$ as factors, where the latter is $H^*$ with
(in our conventions) the opposite
product. This means that a left $D(H)$-module is precisely a vector space $V$
on which $H$ and $H^{*\rm op}$ are represented in a compatible way on the left,
or equivalently on which $H, H^*$ act from the left and right respectively.
Denoting the actions $\la,\ra$, the compatibility condition is \cite{Ma:dou}
\eqn{e34}{\sum <a\o,h\o>h\t\la(v\ra a\t)=\sum (h\o\la v)\ra a\o<a\t,h\t>,\quad
v\in V,\ a\in H^*,\  h\in H.}
The action of $D(H)$ is then $(a\tens h)\la v=(h\la v)\ra a$. Note that a right
$V^*$-module corresponds in the finite-dimensional case to a left $H$-comodule.
So we can equally well think of $V$ as a left $H$-module and $H$-comodule
in a compatible way. This category is then the category of $H$-crossed
`bimodules' studied in \cite{Yet:rep} as well as subsequently by other authors.
It clearly makes sense in this form for any Hopf algebra or bialgebra (not
necessarily finite dimensional).

This $D(H)$ is a quantum group as explained by Drinfeld\cite{Dri} with (in our
conventions) $\CR=\sum (f^a\tens 1)\tens (1\tens e_a)$ where $\{e_a\}$ is a
basis of $H$ and $f^a$ a dual one. Moreover, it is also known that it is
factorizable\cite{ResSem:mat}, so we can apply Proposition~2.2 etc. For
example, the map $Q:D(H)^*\to D(H)$ provided by $\CR_{21}\CR_{12}$ easily comes
out from (\ref{e30}) as
\eqn{e35}{Q(h\tens a)=\sum <a,e_a\t>f^a\tens (\ant e_a\o) h e_a\th}
\eqn{e36}{Q^{-1}(a\tens h)=\sum e_a\o h \ant e_a\th\tens f^a <a,e_a\t>.}
Equally easily, the quantum adjoint action on $D(H)$ comes out as
\eqn{e37}{h\la(a\tens g)=\sum a\t\tens h\t g \ant h_{(4)} <a\o,\ant
h\o><a\th,h\th>}
\eqn{e38}{(b\tens h)\ra a=\sum (\ant^{-1} a\th) b  a\o\tens h\t <a_{(4)},h\o>
<\ant^{-1}a\t,h\th>.}By these actions the braided group $\und {D(H)}$
canonically associated to
$D(H)$ by the construction in (\ref{e3}) lives in the quasitensor category of
$D(H)$-modules. We denote it $B\! D(H)$.

\begin{propos} Let $H$ be a finite-dimensional Hopf algebra. The braided group
$B\! D(H)$ associated to $D(H)$ has the same product (\ref{e30}) but modified
coproduct, inverse antipode and braiding given by
\[ \und\Delta(a\tens h)=\sum a\o\tens e_a\tens f^a\o a\t \ant f^a\th\tens
h\o<f^a\t,h\t>\]
\[ \und\ant^{-1}(a\tens h)=\sum (\ant f^a\o)(\ant a\t) f^a\th\tens e_a
<a\o(\ant f^a\t)\ant a\th,h>\]
\[ \Psi((a\tens h)\tens (b\tens g))=\sum e_a\la(b\tens g)\tens (a\tens h)\ra
f^a.\]
\end{propos}
\proof The braiding is simply from (\ref{e1}) and the known form of $\CR$ for
$D(H)$ (in some examples we can fruitfully evaluate it further). The
braided-coproduct from  (\ref{e3}) comes out as
\align{&&\nqquad\und\Delta (a\tens h)=\sum (a\o\tens h\o)(1\tens \ant e_a)\tens
(a\t\tens h\t)\ra f^a\\
&&=\sum a\o\tens h\o\ant e_a\tens (\ant^{-1} f^a\th)a\t f^a\o\tens h\th
<f^a{}_{(4)},h\t><\ant^{-1} f^a\t,h_{(4)}>\\
&&=\sum a\o\tens h\o e_a\tens f^a\t a\t \ant f^a{}_{(4)}\tens h\th < \ant
f^a\o,h_{(2)}><f^a\th,h_{(4)}>}
where the second equality is from (\ref{e30}) and the third by a change of
basis and dual basis. This gives the expression stated in the proposition
because
$\sum h\o e_a\tens <\ant f^a\o,h\t>f^a\t=\sum e_a\tens \eps(h) f^a$ for all
$h\in H$ (this is easily seen by evaluating on a test function in $H^*$).
Likewise, the inverse braided-antipode is computed from a formula similar to
(\ref{e3}) for $\und\ant$ as
\align{&&\nqquad \und\ant^{-1}(a\tens h)=\sum [\ant^{-1}((a\tens h)\ra
f^a)](1\tens e_a)\\
&&=\sum [\ant^{-1}((\ant^{-1} f^a\th) a f^a\o\tens h\t)](1\tens
e_a)<f^a{}_{(4)},h\o><\ant^{-1} f^a\t,h\th>\\
&&=\sum (1\tens \ant^{-1} h\t)((\ant f^a\o)
(\ant a) f^a\th\tens
e_a)<f^a{}_{(4)},h\o><\ant^{-1} f^a\t,h\th>\\
&&=\sum (\ant f^a\t)(\ant a\t) f^a{}_{(4)}\tens (\ant^{-1} h\t)e_a<a\o
f^a\o,h\o><(\ant f^a\th)(\ant a\th),h\th>\\
&&=\sum (\ant f^a\o)(\ant a\t) f^a{}_{(3)}\tens e_a<a\o,h\o><(\ant f^a\t)
(\ant a\th),h\t>}
as required. For the second equality we used (\ref{e38}), for the third the
(inverse) antipode in $D(H)$ and for fourth the product in $D(H)$, see
(\ref{e30}). For the last equality we used $\sum e_a\ant h\t\tens <
f^a\t,h\o>f^a\o=\sum e_a\tens\eps(h)f^a$ for any $h$. \endproof

{}From Proposition~2.2 we know that this braided group of enveloping algebra
type is also isomorphic to the braided group
$\und{D(H)^*}$ of function algebra type (via $Q$), i.e. $B\! D(H)$ is
self-dual. In our case this is manifest for the product on $B\! D(H)$ is the
same as that of $D(H)$ in (\ref{e30}): an easy computation from (\ref{e30})
gives
its dual (the coproduct on $D(H)^*$ and $\und{D(H)^*}$) as
\eqn{e39}{\Delta_{D(H)^*}(h\tens a)=\sum h\t\tens (\ant f^a\o)a\o f^a\th\tens
e_a\tens a\t <f^a\t,h\o>}
which can be compared with the preceding proposition. Thus the product and
coproduct on $B\! D(H)$ are manifestly isomorphic when compared by dualizing
one of them, i.e. $B\! D(H)$ is in a certain sense `linearized'.
This is a general feature of the braided groups associated to quantum groups,
and allows for them properties usually reserved for $\R^n$. For example, there
is an operation $\CS$ of `quantum Fourier transform' from the braided group to
itself
given by\cite{LyuMa:bra} $\CS=\sum \ant Q^\uo\mu(Q\ut(\ ))$ where $\mu$ is a
suitably normalized left invariant integral and $Q=\CR_{21}\CR_{12}$. Just as
the square of the Fourier transform on $\R^n$ is the parity operator (inversion
on the group), we have $\CS^2=\und\ant^{-1}$\cite{LyuMa:bra}. Moreover, if the
original quantum group is a ribbon Hopf
algebra\cite{ResTur:rib} then there is also an operator $\CT$ given by left
product by the inverse ribbon element, and $(\CS\CT)^3=\lambda\CS^2$ for some
constant $\lambda$. For $B\! D(H)$ the `quantum Fourier transform' is easily
computed from the above as
\eqn{e40*}{\CS (a\tens h)=\sum e_a\tens \ant \mu_1\t\, \mu_2(\ant^{-1} f^a\t
h)<a,f^a\th \mu_1\o \ant^{-1} f^a\o>}
where $\mu_1$ and $\mu_2$ are left integrals on $H^*$ and $H$ respectively,
suitably normalized. For example, the left integral on $H$ is characterized by
$\sum h\o\tens \mu_2(h\t)=1\mu_2(h)$ for all $h\in H$.

Let us note also that the structure of $D(H)$ and $BH(H)$ can also be expressed
fruitfully in terms of $\Hom_k(H,H)$ rather than as we have computed on
$H^*\tens H$. This new form is slightly more general and its structure is a
twisted version of the standard convolution bialgebra on $\Hom_k(H,H)$. To be
explicit, the structure of $D(H)$ and $B\! D(H)$ in these terms is
\eqn{e41*}{(\phi\psi)(g)=\sum \phi(g\t)\t \psi\left((\ant\phi(g\t)\o) g\o
\phi(g\t)\th\right),\  (\Delta\phi)(g\tens h)=
\sum\phi(gh)\o\tens\phi(gh)\t}
\eqn{e42*}{(\ant \phi)(g)=\sum \ant e_a\t<f^a,\phi(e_a\o(\ant^{-1} g)\ant
e_a\th)>}
\eqn{e43*}{ (h\la \phi)(g)=\sum h\t\phi((\ant h\o)g h\th)\ant h_{(4)}}
\eqn{e44*}{(\phi\ra a)(g)=\sum a\left(g\th(\ant^{-1} \phi(g\t)\th)(\ant^{-1}
g\o) \phi(g\t)\o\right)\, \phi(g\t)\t}
\eqn{e45*}{(\und\Delta\phi)(g\tens h)=\sum h\o\phi(g h\t)\t\ant h\th\tens
\phi(g h\t)\o}
\eqn{e46*}{(\und \ant^{-1} \phi)(g)=\sum (\ant g\o)(\ant e_a\t)
g\th<f^a,\phi(e_a\o(\ant^{-1} g\t)\ant e_a\th)>.}
Other structures such as $\Psi$ and $\CS$ are easily computed from those
already computed above in the $H^*\tens H$ form so we leave these to the
reader. The strategy is to replace $h<a,\ >$ (in (\ref{e40*}) for exam
To conclude the general theory we mention that there are plenty of other
algebraic structures naturally living in the
present category of $D(H)$-modules. The following is a version of a theorem of
Radford\cite{Rad:str}, translated into the present context.

\begin{propos} (cf \cite{Rad:str}) Let $H_1{ p\atop{{\longrightarrow\atop
\hookleftarrow}\atop i}}H$ be a Hopf algebra projection (i.e., $p,i$ are Hopf
algebra homomorphisms between two Hopf algebras and $p\circ i=\id$). For
simplicity
we suppose $H$ finite dimensional. Then there is a Hopf algebra $B$ living in
the quasitensor category of $D(H)$-modules such that $B\cocross H\isom H_1$.
Explicitly, $B$ is a subalgebra of $H_1$ and a $D(H)$-module by
\[ B=\{b\in H_1\ |\ \sum b\o\tens p(b\t)=b\tens 1\},\quad h\la b=\sum
i(h\o)b\ant \circ i(h\t),\quad b\ra a=\sum <a,p(b\o)>\tens b\t\]
where $h\in H$ and $a\in H^*$. The braided-coproduct, braided-antipode and
braiding of $B$ are
\[ \und\Delta b=\sum b\o\ant \circ i\circ p(b\t)\tens b\th,\quad \und\ant
b=\sum i\circ p(b\o)\ant b\t,\quad\Psi(b\tens c)=\sum p(b\o)\la c\tens b\t.\]
The structure of $B\cocross H$ is the standard semidirect product by the action
$\la$ of $H$ and the coaction of $H$
corresponding to $\ra$ as stated. The isomorphism $\theta:B\cocross H\to H_1$
is $\theta(b\tens h)=b i(h)$, with inverse $\theta^{-1}(a)=\sum a\o\ant\circ
i\circ p(a\t)\tens p(a\th)$ for $a\in H_1$.
\end{propos}
\proof  The only new part beyond \cite{Rad:str} is the identification of the
`twisted Hopf algebra' $B$ now as a Hopf algebra living in a quasitensor
category, and some  slightly more explicit formulae for its structure. The set
$B$ coincides with the image of the projection $\Pi:H_1\to H_1$ defined by
$\Pi(a)=\sum a\o \ant \circ i\circ p(a\t)$
in \cite{Rad:str}, while the pushed-out left adjoint coaction of $H$ on $B$
then reduces to the left coaction $b\mapsto\sum b\bo\tens b\bt=\sum p(b\o)\tens
b\t$ as used to define $\ra$ in the proposition. Note that the restriction to
$H$ finite-dimensional is avoided if we work throughout with this coaction
rather than $\ra$ as explained above
(this is a reason why comodules are preferred in \cite{Rad:str}). In the
present terms, we obtain an action of $D(H)$.
The braiding from (\ref{e1}) then comes out as $\Psi(b\tens c)=\sum e_a\la
c\tens b\ra f^a=\sum b\bo\la c\tens b\bt$
giving the form shown. The axioms of a Hopf algebra in a quasitensor category
require that $\und \Delta:B\to B\tens B$ is an algebra homomorphism with
respect to the braided tensor product algebra structure on $B\tens B$. Writing
$\und\Delta b=\sum b_{\und{(1)}}\tens b_{\und{(2)}}$, this reads $\und\Delta
(bc)=\sum b_{\und{(1)}}\Psi(b_{\und{(2)}}\tens c_{\und{(1)}})c_{\und{(2)}}=\sum
b_{\und{(1)}}\, (b_{\und{(2)}}\bo\la c_{\und{(1)}})\tens b_{\und{(2)}}\bt
c_{\und{(2)}}
$ which is the condition in \cite{Rad:str}. This, along with the other axioms
of a Hopf algebra in our quasitensor category  can also be easily verified
directly from the formulae stated. Finally, the structure on $B\cocross H$ is
the standard
semidirect product one, $(b\tens h)(c\tens g)=\sum b (h\o\la c)\tens h\t g$ and
the standard semidirect coproduct  by $\ra$ as a coaction. Explicitly, the
coproduct on $B\cocross H$ comes out as $\Delta (b\tens h)=\sum
b_{\und{(1)}}\tens p(b_{\und{(2)}}\o)h\o\tens b_{\und{(2)}}\t\tens h\t$.
Applying $\theta$ to these structures and evaluating further gives $\theta$ as
a Hopf algebra isomorphism. \endproof

We now further compute some of these constructions for an important class of
examples, namely for $D(G)=D(kG)$ where $G$ is a finite group and $kG$ is its
group algebra over a field $k$. Firstly, a $D(G)$-module means a vector space
$V$ on which $G$ and $k(G)$ (functions on $G$ with pointwise product) act. As
is well-known, an action of $k(G)$ simply means a $G$-grading, see for
example\cite{Coh:hop}. Indeed, writing $v\ra a=\sum_g a(g)\beta_g(v)$ for some
operators $\beta_g:V\to V$, the requirement of an action means that
$\beta_g\beta_{g'}=\beta_g\delta_{g,g'}$. Hence $V=\oplus_g V_g$ for
homogeneous subspaces $V_g$ where $a$ acts by $v\ra a=v a(g)$. We say that
$v\in V_g$ has degree $\vert v\vert=g$. Note that $G$ may be non-Abelian. A
$D(G)$-module then means a $G$-graded space on which $G$ also acts, in a
compatible way according to (\ref{e34}). This clearly reduces in our example to
the condition
\eqn{e40}{ \vert g\la v\vert=g\vert v\vert g^{-1},\quad \forall v\in V,\ g\in
G.}
Thus a $D(G)$-module is a $G$-graded $G$-module obeying (\ref{e40}). Note that
$D(G)$ is an ordinary semidirect product (even without appealing to
Proposition~4.1) because $kG$ is cocommutative, so that (\ref{e30}) simplifies.
The braiding from (\ref{e1}) comes out as
\eqn{e41}{\Psi(v\tens w)=\vert v\vert\la w\tens v}
for $v$ homogeneous of degree $\vert v\vert$.

A large class of $D(G)$-modules is provided by the crossed $G$-sets of
Whitehead\cite{Whi:com}. A crossed $G$-set is a set $M$ on which $G$ acts,
together with a map $\del: M\to G$ such that $\del(g\la m)=g(\del m)g^{-1}$ for
all $g\in G, m\in M$. In this case the vector space $kM$ with basis $m\in M$
clearly becomes a $D(G)$-module with $\la$ extended linearly and degree $\vert
m\vert=\del m$. If $M$ is a group one usually demands that $\del$ is a group
homomorphism.
In this case it is easy to see that the group algebra $kM$ is an algebra in the
category of $D(G)$-modules (i.e. the product is $D(G)$-equivariant).  The
braiding (\ref{e41}) is also well known to algebraic topologists. Note that a
further natural demand to make in this context is $(\del m)\la n=mnm^{-1}$ for
all $m,n\in M$\cite{Whi:com}.

Bialgebras or Hopf algebras in this category of $D(G)$-modules obey the usual
axioms after allowing for $\Psi$ in (\ref{e41}) when defining the braided
tensor product algebra structure. We now describe the braided group $B\! D(G)$
associated to $D(G)$. It lives in this category of $D(G)$-modules. Firstly we
adopt a convenient description of $D(G)$ itself. Namely, we identify its
underlying linear space as $k(G)\tens kG=k(G,kG)$, i.e. $kG$-valued functions.
The group-valued functions $\phi_{g,h}=f^g\tens h$ for $g,h\in G$ provide a
basis. Here $\phi_{g,h}(g')=\delta_{g,g'}h$.
We will say that a function $\phi\in k(G,kG)$ is {\em group-like} if the value
at any $g$ is grouplike, $\Delta(\phi(g))=\phi(g)\tens\phi(g)$. They are sums
with each of the basis functions appearing at most once, and can be
thought of as maps from $G\cup 0\to G\cup 0$ where $0$ is adjoined. In this
notation (there are plenty of others) the
quantum group structure of $D(G)$ looks like
\eqn{e42}{(\phi\psi)(g)=\phi(g)\psi\left(\phi(g)^{-1}g\phi(g)\right),\quad
(\Delta\phi)(g,h)=\phi(gh)\tens\phi(gh),\quad
\ant \phi_{g,h}=\phi_{h^{-1}g^{-1}h,h^{-1}}}
These are twisted variants of the usual algebra with point-wise product and
values in $kG$. The antipode looks as stated on one of the basis functions. On
a general group-like function it takes the form $(\ant\phi)(g)=\sum_{h\in
G}h^{-1}\delta_{h,\phi(h g^{-1} h^{-1})}$, which is basically an inversion of
$\phi$ as a map (the resulting function is not in general group-like, however).

\begin{propos} The braided group $B\! D(G)$ associated to $D(G)$ for $G$ a
finite group has the following
structure for group-like functions $\phi,\psi:G\cup 0\to G\cup 0$,
\[ (\und\Delta\phi)(g,h)=h\phi(gh)h^{-1}\tens \phi(gh),\quad (\und
\ant^{-1}\phi)(g)=g^{-1}(\ant \phi)(g)g\]
\[  (h\la\phi)(g)=h\phi(h^{-1}gh)h^{-1},\quad (\phi\ra a)(g)=a([g,\phi(g)])\,
\phi(g)\]
\[(\Psi(\phi\tens\psi))(g,h)=([h,\phi(h)]\la \psi)(g)\tens\phi(h)\]
where $[g,\phi(g)]\equiv g \phi(g)^{-1} g^{-1} \phi(g)$ is the group
commutator. Thus the basis functions $\phi_{g,h}$ are homogeneous of degree
$[g,h]$.
\end{propos}
\proof This, as well as (\ref{e42}), follows more easily from the `twisted
convolution' form in (\ref{e41*})-(\ref{e46*}) by simply dropping the $\uo,\ut$
etc suffices for the coproduct in $H$ (this is the meaning of the group-like
assumption). The original form on $H^*\tens H$ is more useful for the antipode
and braided inverse antipode on the basis functions $\phi_{g,h}$. Another form
for the latter is $\und\ant^{-1}\phi_{g,h}=\phi_{h^{-1}g^{-1}h,h^{-1}[g,h]}$.
\endproof

Finally, as an application we note the quantum Fourier transform operator $\CS$
from \cite{LyuMa:bra}. In fact, $D(G)$ is a ribbon Hopf algebra so that there
is an operator $\CT$ as explained. The inverse ribbon element is simply the
identity map $G\cup 0 \to G\cup 0$ in $D(G)$. The integral on $D(G)$ is the
tensor product one,  $\mu(\phi)=\sum_{g\in G}\delta_{e,\phi(g)}$ (the number of
points in the inverse image of the identity $e$ under $\phi$). From these
observations along with (\ref{e40*}) and a direct computation for the
normalizations, we have
\eqn{e43}{\CS\phi_{g,h}=\phi_{h,h^{-1}g^{-1}h},\quad
(\CT\phi)(g)=g\phi(g),\quad (\CS\CT)^3=\CS^2.}
The action of $\CS$ on group-like functions is $(\CS\phi)(g)=\sum_h
h^{-1}\delta_{g,\phi(g h g^{-1})}$. These expressions seem at first far removed
from ordinary Fourier transforms yet they have similar abstract properties and
moreover,
when the quantum Fourier transform $\CS$ is computed for the quantum
deformations of ordinary Lie groups (such as $\R^n$), it does recover a
deformation of ordinary Fourier transform\cite{LyuMa:bra}. The $D(G)$
example has a different character from these quantum deformations, being a
twisted tensor product of $k(G)=k\hat G$
(in the Abelian case) and $kG$. In this case we see in (\ref{e43}) that that
$\CS$ interchanges the roles of $k(G)$ and $k G$ in $\phi_{g,h}$. In the
Abelian case it is $\CS\phi_{g,h}=\phi_{h,g^{-1}}$. Thus there are some
similarities with
ordinary Fourier transformation in its role of interchanging `position' and
`momentum'.

\end{document}